\documentclass{article}
\usepackage[utf8]{inputenc}
\usepackage[english]{babel}
\usepackage[a4paper,left=1.4in,right=1.4in,top=1.2in,bottom=1.2in,footskip=.25in]{geometry}
\date{May 31, 2019}

\newcommand{\HRule}[1]{\rule{\linewidth}{#1}} 	% Horizontal rule

\makeatletter							% Title
\def\printtitle{						
    {\centering \@title\par}}
\makeatother									

\makeatletter							% Author
\def\printauthor{%					
    {\centering \large \@author}}				
\makeatother

\title{	\normalsize%\textsc{Title page subtitle} 	% Subtitle
			\HRule{0.5pt} \\						% Upper rule
			\LARGE \textbf{\uppercase{Tests of General Relativity and\\ Modified Gravity using Pulsar\\ Timing}} \HRule{2pt}	% Title
		}

\author{
		Cyril Renevey\\
		Supervisor : Philippe Jetzer\\
		UZH and ETH Zürich
}

\usepackage{hyperref}
\hypersetup{colorlinks=true}

\usepackage{array}
\usepackage{tensor}
\usepackage{amsthm}
\usepackage{amsmath}
\usepackage{amssymb}
\usepackage{chngcntr}
\usepackage{commath}
\usepackage{caption}
\usepackage{graphicx}
\usepackage{graphics}
\usepackage{physics}
\usepackage{mathtools}
\usepackage{mhchem}
\usepackage{multirow}
\usepackage{float}

\usepackage{titlesec}
\titleformat*{\paragraph}{\large\bfseries}

\usepackage{multicol}
\newcommand{\half}{\frac{1}{2}}
\newcommand{\alphah}{\hat{\alpha}}

\newcommand{\xih}{\hat{\xi}}

\newcommand{\omegah}{\hat{\omega}}
\newcommand{\etah}{\hat{\eta}}
\newcommand{\gdot}{\dot{G}}
\newcommand{\pdot}{\dot{P}}
\newcommand{\munu}{{\mu\nu}}
\newcommand{\gt}{\tilde{g}}
\newcommand{\LL}{\mathcal{L}}
\newcommand{\EE}{\mathcal{E}}
\newcommand{\ab}{{\alpha \beta}}
\newcommand{\KK}{\mathcal{K}}
\newcommand{\lp}{\left(}
\newcommand{\rp}{\right)}
\begin{document}
\maketitle

\thispagestyle{empty}

\begin{abstract}
    In this report we aim to describe the most stringent tests of the strong equivalence principle, the fundamental principle of General Relativity, using pulsar timing. For this purpose, we first construct the parametrized post-Newtonian and post-Keplerian formalism together with their parameters. Then we constrain the post-Newtonian parameters associated to the violation of the strong equivalence principle with the most stringent tests using pulsar timing techniques. We will in particular see that these bounds are the most constraining limits on a violation of the principle. Finally we will discuss the implication of these results on scalar-tensor theories as well as massive gravity. We will find that the set of possible scalar-tensor theories can be tightly reduce, however in the case of massive gravity, the level of precision is not sufficient to probe the predicted mass of the graviton. 
\end{abstract}
\newpage
\setcounter{page}{1}
%-------------------------------

\section{Introduction}
General Relativity is yet the fundamental theory of gravity and together with the $\Lambda$CDM model, they describe, with great precision, the behaviour of our Universe. However, the Standard Model of Cosmology fails to explain the origin of dark matter and dark energy, which are essential for the theory, and new observations in contradiction with the $\Lambda$CDM model, such as the recent Hubble constant problem, start to emerge. Furthermore, we are yet unable to reconcile gravity and the Standard Model of particle physics at the quantum level. For these reasons, theoretical physicists are working on new gravity theories, including string theory, massive gravity or other theories only bounded by today's experimental data. However, no satisfactory answer to these questions has been found, thus we need more clues to continue investigating on alternative theories of gravity.\\
One of the fundamental property of General relativity is the strong equivalence principle and in order to extend the currently accepted theory of gravity, most of these alternative theories require a violation of the principle. Indeed testing the validity of the strong equivalence principle enables cosmologists to constrain or rule out alternative theories and hopefully to find hints of new physics. Extensive tests of the principle as been conducted, with great precision, in solar system experiments and up to now, observations support the strong equivalence principle. However solar system experiments are restricted to the weak gravitational field regime and we need a way to test General Relativity and the strong equivalence principle in the strong field regime. For this purpose, the study of pulsars is a great help. Indeed these particular neutron stars are very compact objects and one need the realm of strong field gravity to properly describe their behaviour. Using pulsar timing techniques astrophysicists and cosmologists are able to test the validity of General Relativity in one of the most extreme condition, i.e. in compact objects and in their vicinity. 

It is in this context that we write a review of the current results of pulsar timing tests of General Relativity and modified gravity theories. Since the discovery of the Hulse-Taylor binary pulsar in 1974, numerous experiments has been conducted in order to test the strong equivalence principle and modified gravity theories in the strong gravitational field regime. Our goal is to report and describe the most stringent tests of the principle using pulsar timing techniques and to use the current observations in order to constrain scalar-tensor theories and massive gravity. After a small introduction on pulsars and pulsar timing in section \ref{section_pulsar}, we will describe, in section \ref{section_PPN} and \ref{section_PPK}, the theoretical framework required to understand the different tests of the strong equivalence principle, namely the parametrized post-Newtonian and post-Keplerian formalism together with their parameters. Section \ref{section_SEP} constitutes the core content of this report, we will talk about the different possible tests of the strong equivalence principle and mention the most stringent results that has been found. Using these results, we will constrain, in section \ref{section_ST}, the simplest class of modified gravity theories, i.e. scalar-tensor theories. In section \ref{section_MG} we will finally discuss the case of massive gravity and mention the constraint, from pulsar observations, on the mass of the graviton.
%---------------------------

\section{Properties of Pulsars}\label{section_pulsar}
Radio pulsars have been confirmed \cite{stairs2003testing} to be rapidly rotating neutron stars with a very strong magnetic field, which can accelerate particles to very high energies. Moreover if the magnetic dipole is not aligned with the rotation axis, the pulsar can emit particles and electromagnetic waves with a precise frequency, in the direction of the magnetic dipole. These objects can be found alone or in a binary system with a white dwarf or an other neutron star companion, for example. In particular the pulsar J0337+1715 has two white dwarf companions \cite{ransom2014millisecond}, which form together a triple system, and this system is particularly interesting to test the weak equivalence principle, as we will see later on.

\subsection{Pulsar's Observation and Timing}

The typical frequencies $f_0$ of the emitted photons are of the order of a few hundred MHz, however at these frequencies, the ionized interstellar medium can alter the sharp picks of the emission and reduce the precision of the measurements. Moreover the effect of scintillation due to interference between different rays following different paths can also alter the measurements. One can reduce these effects with longer time allocations and by observing higher frequencies range as the alterations decrease roughly as $f_0^{-4}$ \cite{stairs2003testing}.

In order to have a good estimation of the time of emission $t$ of the pulsar, one need to take into account, from the observed time of arrival $\tau$, different effects : the Roemer delay $\Delta_{R\odot}$ for the classical travel time through earth orbit, the Shapiro delay $\Delta_{S\odot}$ which accounts for the delay of the signal due to the Sun gravitational field, the Einstein delay $\Delta_{E\odot}$ from time dilation due to the masses in the solar system and the dispersion delay $D/f^2$, where $D=\textrm{DM}/2.41\cdot 10^{-4}$ Hz comes from the pulsar's dispersion measure and $f$ is the observed frequency. If the pulsar is part of a binary system, similar delays $\Delta_E$, $\Delta_S$ and $\Delta_R$, for the system itself, need to be be taken into account. This correction can be summarized with the following formula \cite{stairs2003testing} : 
\begin{align}
    t=\tau-D/f^2+\Delta_{R\odot}+\Delta_{E\odot}-\Delta_{S\odot}-\Delta_{R}-\Delta_{E}-\Delta_{S}.\label{eq_timingformula}
\end{align}
We will describe more in detail the parameters $\Delta_{R}$, $\Delta_{E}$ and $\Delta_{S}$ in the section \ref{PPK_MG} for general alternative theories.
\subsection{Use of Pulsars}
Because of their very precise rotation rate, Pulsars are considered as high precision relativistic clocks and thus one of the best objects to study General Relativity. Using pulsar timing techniques one can measure the Keplerian and post-Keplerian parameters, described in table \ref{table_PPK}, of pulsars and binary systems with very high precision. Furthermore, Pulsars are very compact objects contrary to the celestial objects in our Solar system, and one can easily observe or test higher order phenomenon in post-Newtonian approximation. For example the mean periastron advance of the famous pulsar system PSR B1913+16 is $\ev{\dot{\omega}}=4.23^{\circ}$ /year \cite{will2018theory}, which is very large compared to the perihelion advance of Mercury, $\dot{\omega}\cong 43''$/century. Pulsars can also have a very short rotation period, of the order of milliseconds, and such a fast rotation will have different effect on its motion depending on the theory we are considering. Indeed the spin precession might differ for other theories than General Relativity, which is an effect we can measure and test, as we will see in chapter \ref{section_SEP}. When the pulsar is accompanied by an other celestial object, the compactness of the former will have post-Newtonian effects on the orbital motion. Here again these effects are theory dependent and by measuring the orbital parameters such as the orbital period or the eccentricity, one can test the validity of General Relativity.

%----------------------------------

\section{Parametrized Post-Newtonian Formalism}\label{section_PPN}
We first start our discussion by describing the weak field limit using the post-Newtonian approximation and the PPN parameters. Even though pulsar observations have a bigger impact in strong gravity astronomy due to their compactness and short orbital periods for binary systems, they had a significant impact on limiting different gravity theories and violation of the different equivalence principles in the weak field sector, as we will see in the sections \ref{section_SEP} and \ref{section_ST}. We announce already here that throughout the report, otherwise mentioned, we use the convention $c=G=1$ to match the notation of most of the important sources and for the simplicity of the equations.

\subsection{The PPN Formalism and its Parameters}\label{section_PPNF}

To build the PPN formalism and write a general PPN metric, we start by enumerating the required assumptions :

\begin{enumerate}
    \item Weak gravitational field and small velocities,
    \item matter is described by perfect fluids,
    \item metric should be of Newtonian or post-Newtonian order and dimensionless,
    \item the metric should be assymptotically Minkowskian,
    \item the metric corrections $h_{00}$, $h_{0i}$ and $h_{ij}$ should transform under rotation as scalar, vector and tensor respectively.
\end{enumerate}
Using these assumptions and keeping only post-Newtonian terms in the metric that appear in different gravitational theories, one can write a general metric as \cite{will2018theory} :
\begin{subequations}
\label{eq_PPNmetric}
\begin{align}
    g_{00}&=-1+2U+2(\psi-\beta U^2)+(1-\half \alpha_1+\alpha_2+2\xi)\ddot{X}+\Phi^{PF}_{Harm}+O(\epsilon^3),\\
    g_{0j}&=-\half(4(1+\gamma)+\alpha_1)V_j-\frac{1}{4}\alpha_1 X_{,0j}+\Phi^{PF}_{j,Harm}+O(\epsilon^{5/2}),\\
    g_{jk}&=(1+2\gamma U)\delta_{jk}+O(\epsilon^2),
\end{align}
\end{subequations}
where
\begin{align}
    \psi :=&\half(2\gamma+1-2\xi)\Phi_1-(2\beta-1-\xi)\Phi_2+\Phi_3+(3\gamma-2\xi)\Phi_4+\xi\Phi_6-\xi \Phi_W.
\end{align}
$U$, $V_j$, $X$, $\Phi_1$, $\Phi_2$, $\Phi_3$, $\Phi_4$, $\Phi_6$, $\Phi_W$, $\Phi^{PF}_{Harm}$ and $\Phi^{PF}_{j,Harm}$ are different potentials that can be found in Will 2018 \cite{will2018theory} and $\epsilon\sim v^2$ is the small parameter of the expansion. Note that we are working in the harmonic gauge, i.e. $[(1-(1-\gamma)U\sqrt{-g}g^\munu]_{,\nu}=0$. For more detailed derivations see Will 2018 (\cite{will2018theory}).  The last unmentioned variables are the PPN parameters and have been chosen in such a way that they represent different physical interpretations. Different gravity theories will have different requirements for these parameters and thus it is possible to test different theories by the observation of the PPN parameters. A summary of their interpretations and predicted value for different theories is given in the table \ref{table_PPN}.

\begin{table}
\centering
\begin{tabular}{|c|l|c|c|c|}
\hline
Parameters  & Physical meaning                                                                                             & \multicolumn{1}{l|}{\begin{tabular}[c]{@{}l@{}}GR \\ prediction\end{tabular}} & \multicolumn{1}{l|}{\begin{tabular}[c]{@{}l@{}}Scalar-tensor \\ prediction\end{tabular}} & \multicolumn{1}{l|}{\begin{tabular}[c]{@{}l@{}}TeVeS \\ prediction\end{tabular}} \\ \hline\hline
$\gamma$   & \begin{tabular}[c]{@{}l@{}}Space curvature produced \\ by unit rest mass\end{tabular}                        & $1$                                                                           & $\frac{1+\omega_0}{2+\omega_0}$                                                              & 1                                                                                \\ \hline
$\beta$    & \begin{tabular}[c]{@{}l@{}}Non-linearity in superposition \\ law for gravity\end{tabular}                    & $1$                                                                           & $1+\frac{\lambda}{4+2\omega_0}$                                                            & 1                                                                                \\ \hline
$\xi$      & Preferred-location effects                                                                                   & $0$                                                                           & $0$                                                                                      & $0$                                                                              \\ \hline
$\alpha_1$ & Preferred-frame effects                                                                                      & $0$                                                                           & $0$                                                                                      & $\alpha_1'$                                                                      \\ \hline
$\alpha_2$ & Preferred-frame effects                                                                                      & $0$                                                                           & $0$                                                                                      & $\alpha_2'$                                                                      \\ \hline
$\alpha_3$ & \begin{tabular}[c]{@{}l@{}}Preferred-frame effects and \\ non-conservation of \\ total momentum\end{tabular} & $0$                                                                           & $0$                                                                                      & 0                                                                                \\ \hline
$\zeta_i$  & \begin{tabular}[c]{@{}l@{}}Non-conservation of \\ total momentum\end{tabular}                                & $0$                                                                           & $0$                                                                                      & 0                                                                                \\ \hline
\end{tabular}
\captionof{table}{List of the PPN parameters with their physical meaning and their prediction for GR, scalar-tensor and TeVeS theories \cite{will2014confrontation}. $i=1,...,4$ for $\zeta_i$, $\lambda$ and $\omega_0$ are variables for the scalar-tensor theories, described in section \ref{sect_PPNST}, $\alpha_1'$ and $\alpha_2'$ are highly complex functions.\label{table_PPN}}
\end{table}

It is important to note however that for the case of compact objects, the post-Newtonian formalism is not completely satisfactory and we need to consider analogous parameters, $\hat{\gamma}$, $\hat{\beta}$, $\hat{\xi}$, $\hat{\alpha}_j$ and $\hat{\zeta}_i$, with $j=1,2,3$ and $i=1,...,4$, where the "hat" on the PPN parameters means that we consider the strong field analog. Even though the comparison between the strong field and weak field parameters is not straight forward, we usually expect that strong field corrections of the alternative gravity theories would increase the deviation of a strong field PPN parameter away from its GR value, thus a bound on the strong field parameters can be reasonably translated to a bound on the weak field PPN parameters.

\subsection{Post-Newtonian Parameters for General Relativity}
Our goal is now to find the PPN parameters for the case of General Relativity. As we can observe in the equations \eqref{eq_PPNmetric}, the lowest order in the post-Newtonian approximation (1PN) is for $g_{00}\sim O(\epsilon^2)$, $g_{0i}\sim O(\epsilon^{3/2})$ and $g_{ij}\sim O(\epsilon)$. Furthermore, at this order the gravitational radiation is neglected and thus the system is time reversal. This property of the system is guaranteed if the metric elements $g_{00}$ and $g_{ij}$ are even in $\epsilon$, whereas $g_{0i}$ are odd. Using these arguments and similar ones for the energy-momentum tensor $T_\munu$ we can write the field equations for 1PN corrections to the metric in the de Donder gauge $(\sqrt{-g}g^\munu)_{,\nu}=0$ \cite{Weinberg1972gravitation}
\begin{subequations}
\label{eq_1PNeqt}
\begin{align}
    \Delta (g_{00}^{(1)})=& -8\pi T^{00}_{(0)}\\
    \Delta (g_{ij}^{(1)})=& -8\pi\delta_{ij}T^{00}_{(0)}\\
    \Delta (g_{0i}^{(3/2)})=&\ 16\pi T^{0i}_{(1/2)}\\
    \Delta (g_{00}^{(2)})=&\ \partial_0^2(g_{00}^{(1)}+g_{ij}^{(1)}\partial_i\partial_j(g_{00}^{(1)})-\partial_i(g_{00}^{(1)})\partial_i(g_{00}^{(1)})\nonumber\\&-8\pi\left(T^{00}_{(1)}+T^{ii}_{(1)}-2g_{00}^{(1)}T^{00}_{(0)}\right),
\end{align}
\end{subequations}
where $g_\munu^{(n)},T^\munu_{(n)}\sim O(\epsilon^n)$ and the sum over repeated indices is implicit. Using these equations, one can solve for the different orders of the metric and we obtain the expression for $g_\munu$ at 1PN \cite{Weinberg1972gravitation}
\begin{subequations}
\label{eq_PPNmetricGR}
\begin{align}
    g_{00}&=-1+2U-2(U^2-\psi)+O(\epsilon^3)\\
    g_{0i}&=-4V_i+O(\epsilon^{5/2})\\
    g_{ij}&=\delta_{ij}+2U\delta_{ij}+O(\epsilon^2),
\end{align}
\end{subequations}
where 
\begin{align}
U&=\int\dd^3x'\frac{T^{00}_{(0)}(\vb{x'},t)}{\abs*{\vb{x}-\vb{x'}}},\quad V_i=\int\dd^3x'\frac{T^{0i}_{(1/2)}(\vb{x'},t)}{\abs*{\vb{x}-\vb{x'}}}\nonumber\\
\textrm{and}\quad \psi&=\int\frac{\dd^3x'}{\abs*{\vb{x}-\vb{x'}}}\left(\frac{-1}{4\pi}\partial_0^2U+ T^{00}_{(1)}(\vb{x'},t)+T^{ii}_{(1)}(\vb{x'},t)\right).\label{eq_psiPPNGR}
\end{align}
To connect this result to the general case \eqref{eq_PPNmetric}, we need to write \eqref{eq_psiPPNGR} in terms of the potentials $\Phi_i$, $i=1,...,4$. If we want to follow the notation of Will 2018 \cite{will2018theory}, we need to work with the energy-momentum tensor for perfect fluids, i.e. $T^\munu=(\rho+\rho\Pi+p)u^\mu u^\nu+pg^\munu$. Since $p/\rho,\Pi,v^2,U\sim O(\epsilon)$, we can write the first few orders of $T^\munu$ \cite{will2018theory}
\begin{subequations}
\begin{align}
    T^{00}&=\rho(1+\Pi+\half v^2-U)+O(\epsilon^{3/2}),\\
    T^{ij}&=\rho v^iv^j+p\delta^{ij}+O(\epsilon^{3/2}).
\end{align}
\end{subequations}
From here we can rewrite \eqref{eq_psiPPNGR} as
\begin{align}
    \psi&=\int\frac{\dd^3x'}{\abs*{\vb{x}-\vb{x'}}}\left(\frac{-1}{4\pi}\partial_0^2U+\rho'\Pi'+\half\rho' v'^2-\rho'U'+\rho'v'^2+3p'\right)\nonumber\\
    &=\ddot{X}+\Phi_3+\frac{3}{2}\Phi_1-\Phi_2+3\Phi_4,\label{eq_psiPPNGR2}
\end{align}
where we write $f':=f(\vb{x'},t)$, for $f=v,\rho,U,\Pi,p$. If we now substitute \eqref{eq_psiPPNGR2} into \eqref{eq_PPNmetricGR} and compare with \eqref{eq_PPNmetric}, we can deduce the PPN parameters for General Relativity as mentioned in table \ref{table_PPN}. Note that to get the expressions for $\zeta_i$, $i=1,...,4$ we need to work in a different gauge, namely the standard PPN gauge.

%-----------------------------------------

\section{Keplerian and Post-Keplerian Parameters}\label{section_PPK}
We now present the parametrized post-Keplerian formalism. Contrary to the PPN formalism, the PPK approach is a general framework in the strong field regime and is described by the PPK parameters, a set of variables that can be measured in a theory-independent way. We present, in the table \ref{table_PPK}, the Keplerian and post-Keplerian parameters contained in pulsar timing data. 
\begin{table}
\centering
\begin{tabular}{|lccc|}
\hline
\multicolumn{2}{|c|}{Pulsar parameters}                  & \multicolumn{2}{c|}{Keplerian parameters}                   \\ \hline
Right ascention       & \multicolumn{1}{c|}{$\alpha$}    & \multicolumn{1}{l}{Projected semi-major axis}  & $x= a_1s$     \\
Declination           & \multicolumn{1}{c|}{$\delta$}    & \multicolumn{1}{l}{Eccentricity}               & $e$        \\
Pulsar period         & \multicolumn{1}{c|}{$P_p$}       & \multicolumn{1}{l}{Orbital period}             & $P_b$      \\
Derivative of period  & \multicolumn{1}{c|}{$\dot{P}_p$} & \multicolumn{1}{l}{Longitude of periastron}    & $\omega_0$ \\
\multicolumn{1}{|c}{} & \multicolumn{1}{c|}{}            & \multicolumn{1}{l}{Time of periastron passage} & $T_0$      \\ \hline\hline
\multicolumn{4}{|c|}{Post-Keplerian parameters}                                                                        \\ \hline
\multicolumn{2}{|l}{Mean rate of periastron advance}    & \multicolumn{2}{c|}{$\ev{\dot{\omega}}$}                    \\
\multicolumn{2}{|l}{Redshift / Time dilation}           & \multicolumn{2}{c|}{$\gamma'$}                              \\
\multicolumn{2}{|l}{Derivation of orbital period}       & \multicolumn{2}{c|}{$\dot{P}_b$}                            \\
\multicolumn{2}{|l}{Range of Shapiro delay}             & \multicolumn{2}{c|}{$r$}                                    \\
\multicolumn{2}{|l}{Shape of Shapiro delay}             & \multicolumn{2}{c|}{$s=\sin i$}                                    \\ \hline
\end{tabular}
\captionof{table}{Summary of the Keplerian and post-Keplerian parameters. Different theories of gravity will predict different values for the post-Keplerian parameters. The parameter $i$ represents the angle between the orbital plane and the plane of the sky, perpendicular to the line of sight and $a_1$ is the semi-major axis of the pulsar.(Table from Will 2018 \cite{will2018theory}).\label{table_PPK}}
\end{table}
First derived by Damour and Taylor 1992 \cite{damour1992strong}, the goal of this formalism is to be able to test General Relativity and other gravity theories by direct observation of the parameters presented in the table \ref{table_PPK}. Different theories will predict different values for post-Keplerian parameters and one can directly constraint or rule out theories by observing the PPK parameters for relativistic system such as binary pulsars.

\subsection{Post-Keplerian Parameters For Alternative Theories}\label{PPK_MG}
We now follow the development of Will 2018 \cite{will2018theory} to describe the post-Keplerian parameters for alternative theories. We first need to define a useful coordinate system : we assume that in the vicinity of the binary system, composed of a pulsar of mass $m_1$ and a companion of mass $m_2$, the metric is post-Newtonian, asymptotically Minkowskian and the origin is at the center of mass of the binary system. The time of the observer is denoted $t$ whereas the proper time of the pulsar is represented by $\tau$. We then define the eccentric anomaly $u$ related to the observer time as 
\begin{align}
    u-e\sin u=\frac{2\pi}{P_b}(t-T_0),
\end{align}
where $e$, $P_b$ and $T_0$ are defined in table \ref{table_PPK}. In short, $u$ tells us where is the pulsar on its orbit compare to the periastron. In particular, using the definition of $u$ we can express the distance between the two objects as $r=a_1(1-e\cos u)$, where $a_1$ is the semimajor axis of the pulsar. Furthermore we define the $N^\textrm{th}$ rotation of the pulsar in terms of the rotation frequency $\nu(\tau)$ :
\begin{align}
    N(\tau)=N(0)+\nu\tau+\half\dot{\nu}(0)\tau^2+O(\tau^3).\label{pulse}
\end{align}

To take into account the corrections for compact objects, we use the modified Einstein-Infeld-Hoffmann (EIH) metric (see chapter 10 of Will 2018 \cite{will2018theory}) in order to relate the proper time of the pulsar with the time of a far away observer 
\begin{align}
    \dd\tau^2&=g_{\mu\nu}\dd x^\mu\dd x^\nu=(1-2\alpha_2^*\frac{m_2}{r})\dd t^2-(1+2\gamma_2^*\frac{m_2}{r})\dd \vb{r}^2\nonumber\\
    &=\dd t^2\left(1-2\frac{m_2}{r}(\alpha_2^*+\gamma_2^*v_1^2)-v_1^2\right)\nonumber\\
    \implies \dd \tau&\cong \dd t(1-\alpha_2^*\frac{m_2}{r}-v_1^2),
\end{align}
where $v_1$ is the speed of the pulsar relative to the center of mass of the system, $\alpha_2^*$ and $\gamma_2^*$ are functions of the parameters of a particular theory. We took the first order expansion of the square root and neglected $\gamma_2^*v_1^2$ as compared to $\alpha_2^*$. We also dropped the constant contribution of the pulsar gravitational effects between the pulsar center of mass and the emission point : $\alpha_1^*m_1/\abs{\vb{x}_{em}-\vb{x}_1}$. One can then replace the speed with $v_1^2=\mathcal{G}(m_2^2/m)(2/r-1/a)$, where $m=m_1+m_2$, $a$ is the semi-major axis between the two objects and $\mathcal{G}$ is an EIH parameter, such that the modified EIH equations of motion are $\vb{a}=-\mathcal{G}m\vb{x}/r^3$. We finally integrate between the time of emission $t_e$ and $t=\tau=0$ to obtain
\begin{align}
    \tau\cong t_e-\frac{m_2}{a}(\alpha_2^*+\mathcal{G}\frac{m_2}{m})\frac{P_b}{2\pi}e\sin u,\label{proper}
\end{align}
where we absorbed the constant factor multiplying $t_e$ into the definition of $\nu$ and dropped the constant terms. After its emission the signal will follow a geodesic to arrive at the barycenter of the solar system $\vb{x}_\odot(t_{a})$  at time $t_a$. To compute its true travel time $t_a-t_e$ we can decompose it into the classical travel time and the Shapiro delay due to the companion's gravitational field
\begin{align}
    t_a-t_e=\abs{\vb{x}_\odot(t_a)-\vb{x}_1(t_e)}+(\alpha_2^*+\gamma_2^*)m_2\log\left(\frac{2r_\odot(t_a)}{r(t_e)+\vb{n}\cdot \vb{x}(t_e)}\right),\label{tate}
\end{align}
where $\vb{x}$ is the vector between the pulsar and the companion and $\vb{n}$ is the normal vector of $\vb{x}_\odot$. We don't consider the delay due to the pulsar gravitational field, because it has a constant effect on the arrival time. We now assume $r_\odot\gg r\implies \abs{\vb{x}_\odot(t_a)-\vb{x}_1(t_e)}\cong r_\odot(t_a)-\vb{x}_1(t_e)\cdot\vb{n}$ and we can reset the arrival time as $t_a\rightarrow t_a'=t_a-r_\odot$. Due to this resetting, $t_a'-t_e$ is small and one can expand $\vb{x}_1(t_e)$ around $t_a'$, which gives
\begin{align}
    \vb{x}_1(t_e)\cong \vb{x}_1(t_a')+\vb{v}_1(t_a')(t_e-t_a')\cong \vb{x}_1(t_a')+\vb{v}_1(t_a')(\vb{x}_1(t_a')\cdot \vb{n}).\label{eqx1ta}
\end{align}
We can now substitute \eqref{eqx1ta} into \eqref{tate} and the Shapiro delay is already at first order, so one can send $t_e\rightarrow t_a'$, to finally get
\begin{align}
    t_e=t_a'+(1+\vb{v}_1(t_a')\cdot \vb{n})(\vb{x}_1(t_a')\cdot\vb{n})+\Delta_S(t_a'), \label{te}
\end{align}
where $\Delta_S(t_a')$ is the Shapiro delay. By combining equations \eqref{pulse}, \eqref{proper} and \eqref{te} we can relate the pulse number and the arrival time. We can further include a dependence on the gravitational field of the gravitational constant as
\begin{align}
    G=G_0(1-\eta_2^*\frac{m_2}{r}),
\end{align}
where $\eta_2^*$ depends on the theory. This effect will vary the rotation rate of the pulsar according to
\begin{align}
    \frac{\Delta \nu}{\nu}=-\kappa \eta_2^*\frac{m_2}{r},
\end{align}
with $\kappa$ the sensitivity of the moment of inertia to a change in $G$. Finally we can relate this derivation to the timing formula \eqref{eq_timingformula}, without the correction for the solar system gravitational effect 
\begin{align}
    t_a' = \tau-\Delta_R(u)-\Delta_E(u)-\Delta_S(u),
\end{align}
where $\tau=N/\nu$ and
\begin{subequations}

\begin{align}
    \Delta_R(u)&=xF(e,\omega,u)(1+x\dot{F}(e,\omega,u)),\\
    \Delta_E(u)&=\gamma'\sin u,\\
    \Delta_S(u)&=-2r\log\left(1-e\cos u-sF(e,\omega,u)\right),
\end{align}
\end{subequations}
where
\begin{subequations}
\label{EqPPK}
\begin{align}
    F(e,\omega,u)&=\sin\omega(\cos u -e)+\cos\omega(1-e^2)^{1/2}\sin u,\\
    x&=a_1s=\frac{m_2}{m}as,\\
    \gamma'&=em_2\left(\frac{P_b}{2\pi\mathcal{G}m}\right)^{1/3}(\alpha_2^*+\mathcal{G}\frac{m_2}{m}+\kappa\eta_2^*),\\
    r&=\half(\alpha_2^*-\gamma_2^*)m_2,\\
    s&=\sin i,
\end{align}
\end{subequations}
and $\omega=\omega_0+\ev{\dot{\omega}}(t-t_0)+...$ is the periastron angle. $i$ is the angle between the plane of sky, which is the plane perpendicular to the line of sight, and the orbital plane. We do not derive the expression for $\ev{\dot{\omega}}$ and $\dot{P}_b$ here, a complete development for these PPK parameters is done in chapter 10 and 11, respectively, of Will 2018 \cite{will2018theory}.

We are now equipped with a powerful tool, by direct observation of the pulsar timing we can deduce the post-Keplerian parameters and because each theory of gravity predicts different expressions for $\alpha_2^*$, $\gamma_2^*$, $\eta_2^*$ and $\kappa$, one can test their plausibility. In the equations \eqref{EqPPK}, we can see that we have a dependence of the PPK parameters on the mass of the pulsar and its companion. By observing the value of a specific PPK parameter, one can deduce a curve of possible masses in an $m_1$-$m_2$ diagram and doing so for several parameters gives us several curves, which should intersect at a single point. Further description of this test will be done for General Relativity in the section \ref{PPKGR}.

\subsection{PPK Parameters for General Relativity and the Mass-Mass Diagrams}\label{PPKGR}
It is interesting to first test the dominant theory of gravity, General Relativity, using the PPK parameters. For GR, the theory dependent parameters are given by $\alpha_2^*=\gamma_2^*=\mathcal{G}=1$ and $\eta_2^*=0$. Using these predictions, one can rewrite the PPK parameters \eqref{EqPPK}, in the context of GR, where we add for completeness the expression for $\ev{\dot{\omega}}$ and $\dot{P}_b$ \cite{will2018theory}
\begin{subequations}
\label{GRPPK}
\begin{align}
    \gamma'&=em_2\left(\frac{P_b}{2\pi m}\right)^{1/3}(1+\frac{m_2}{m}),\\
    r&=m_2,\\
    s&=\sin i=x\left(\frac{2\pi}{P_b}\right)\frac{m^{2/3}}{m_2},\\
    \ev{\dot{\omega}}&=\frac{6\pi}{P_b(1-e^2)}\left(\frac{2\pi m}{P_b}\right)^{2/3},\\
    \dot{P_b}&=-\frac{192\pi}{5}\left(\frac{2\pi M}{P_b}\right)^{5/3}f(e),
\end{align}
\end{subequations}
where
\begin{align}
    M&=m\left(\frac{m_1m_2}{m^2}\right)^{3/5}\qquad \textrm{the chirp mass},\nonumber\\
    f(e)&=(1-e^2)^{-7/2}\left( 1+\frac{73}{24}e^2+\frac{37}{96}e^4 \right).\nonumber
\end{align}
As previously mentioned, by direct observation of the PPK parameters one can deduce five independent curves on a mass-mass diagram using the equations \eqref{GRPPK}. Because this test involves only two unknown variables, the mass of the pulsar $m_1$ and the mass of its companion $m_2$, but five independent equations, it is a very powerful test of the theory. One of the most stringent result of this test has been conducted by Kramer et al 2006 \cite{kramer2006tests} on the binary system PSR J0737-3039A/B. This system is composed of two pulsars, which makes it very interesting to study and gives very precise measurements of the Keplerian and post-Keplerian parameters. The mass-mass diagram for the system PSR J0737-3039A/B, assuming General Relativity is exposed in figure \ref{MMdiagram}.

\begin{center}
    \includegraphics[width=0.6\textwidth]{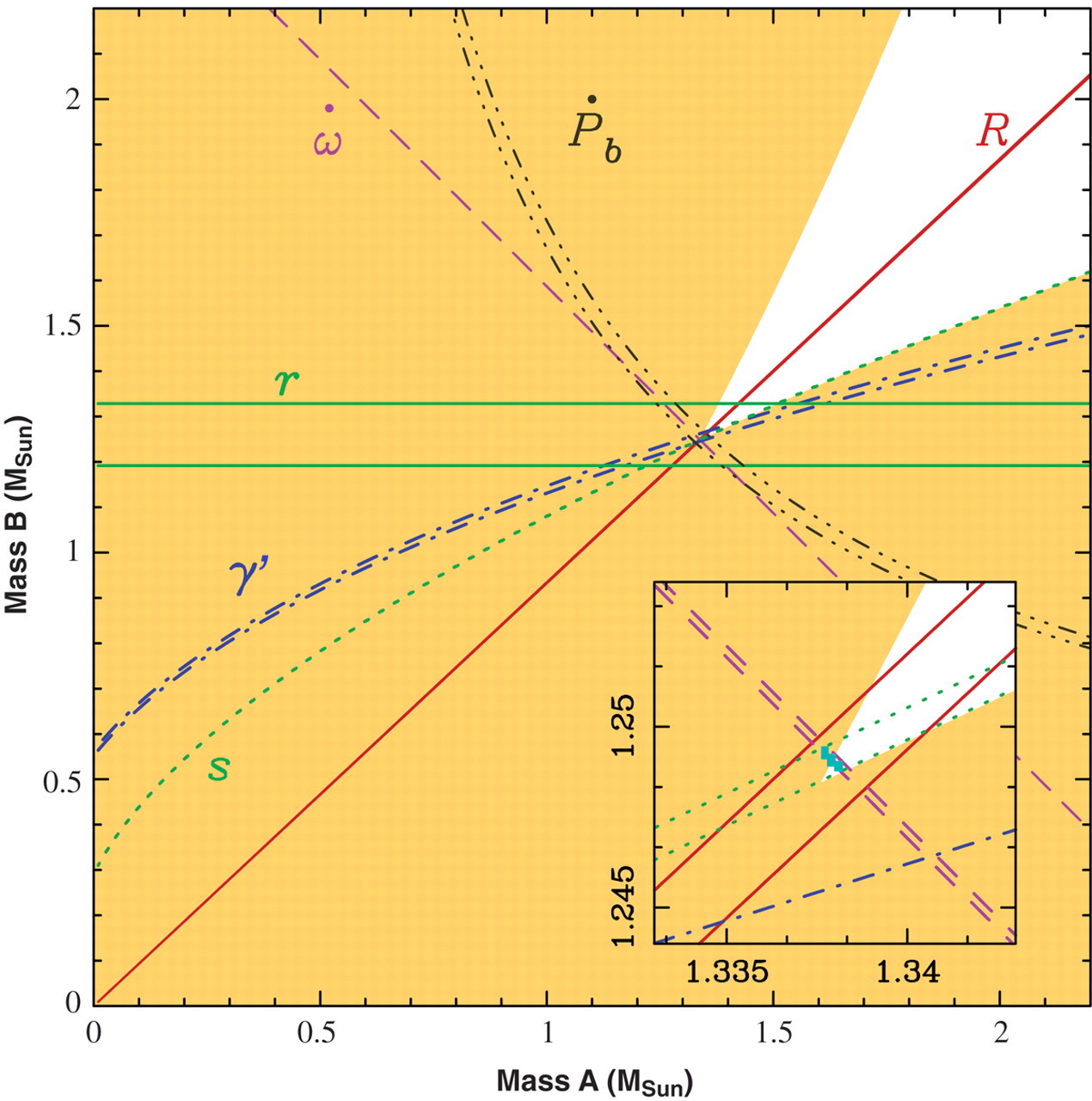}
    \captionof{figure}{Mass-mass diagram of the binary system  PSR J0737-3039A/B, assuming GR. $\dot{\omega}$, $r$, $s$, $\dot{P}_b$ and $\gamma'$ are the PPK parameters as described in table \ref{table_PPK}. $R$ represents the mass ratio derived from the semi-major axis measurements. Each curve is drawn with two lines representing its confidence interval and the turquoise shade in the zoomed image is the superposition of all the confidence intervals. (Image from Kramer et al 2006 \cite{kramer2006tests}).\label{MMdiagram}}
\end{center}

As we can see in figure \ref{MMdiagram}, General Relativity is compatible with the measurement of the PPK parameters. Indeed there is a small area for which all the curves intersect within their confidence interval. Furthermore, from this diagram, one can deduce the masses of the two objects as predicted by General Relativity. One can repeat this test for other theories of gravity, which have different predictions for the PPK parameters, and constraint the parameters of the alternative theory or rule out theories which do not fit the observations.

%-----------------------------------------

\section{Tests of the Strong Equivalence Principle}\label{section_SEP}
With the help of the PPN formalism, it is possible to test the strong equivalence principle (SEP) using the PPN parameters summarized in the table \ref{table_PPN}. The studies of single pulsars, binary systems and the recently discovered triple-star system J0337+1715, were able to set upper limits on different PPN parameters related to the strong equivalence principle. In this section we will describe the most recent tests which were able to set the best upper limits on the strong field parameters $\hat{\alpha}_1$, $\hat{\alpha}_2$, $\hat{\alpha}_3$ and $\hat{\xi}$, as well as explaining the connection between these parameters and the strong equivalence principle. The latter states that \cite{straumann2012general, will2018theory}

\begin{enumerate}
    \item In an external gravitational field, the motion of test bodies and self-gravitating bodies is independent of its mass and composition. This statement is an extension of the weak equivalence principle, called the gravitational weak equivalence principle (GWEP).
    \item There exists no preferred reference frame, i.e. Lorentz invariance.
    \item There is no preferred position in the universe, i.e. position invariance.
\end{enumerate}

The variation of Newton's constant is also considered as a violation of the SEP, more precisely a violation of position invariance. We will then discuss the relevant tests on the variation of Newton's constant. Most of the modified gravity theories include violation of the strong equivalence principle, thus testing its validity is of capital importance in the research of a new candidate for general relativity. The recent discovery of the triple system sets an upper limit on the first principle using the strong field approximation of the Nordtvedt effect, an other result that we will study in this section.

\subsection{Limit on $\vb*{\hat{\alpha}_1}$ using Orbit Eccentricity}

As we have seen in table \ref{table_PPN}, $\alphah_1$ and $\alphah_2$ are associated to preferred-frame effect, i.e. the second principle we introduced, thus a non-zero value for these parameters would be translated as a violation of the SEP. Starting with the first parameter $\alphah_1$, a non-zero value for the latter means that the velocity $\vb{w}$ of the center of mass of a binary system relative to a preferred frame, such as the cosmic microwave background (CMB), will have an impact on the orbital evolution of the binary system. Damour and Esposito-Farèse \cite{damour1992testing} have shown that for a system with low eccentricity $e$  composed of a pulsar and a light companion such as a white dwarf, the eccentricity vector $\vb{e}$, can be written as \cite{stairs2003testing} :
\begin{align}
    \vb{e}(t)&=\vb{e_R}(t)+\vb{e_F},\label{e}\\
    \abs{\vb{e_F}}&=\frac{1}{12}\alphah_1\abs{\frac{m_1-m_2}{m_1+m_2}}\frac{\abs{w_\perp}}{[ (m_1+m_2)2\pi/P_b]^{1/3}},\label{ef_norm}
\end{align}
where $\vb{e_R}$ is the change of eccentricity due to the advance of periastron $\dot{\omega}$, $\vb{e_F}$ is a constant forced eccentricity due to a non-zero $\alphah_1$, $m_1$ is the mass of the pulsar, $m_2$ the mass of the companion, $w_\perp$ the projection of $\vb{w}$ on the orbital plane and $P_b$ is the orbital period. We call $\theta (t)$ the angle between $\vb{e_R}(t)$ and $\vb{e_F}$. To assume a uniformly distributed initial $\theta$, the system should be old enough to have let $\vb{e_R}(t)$ completely rotate several times. If the observation period $T$ is long enough, the change in $\theta$ will be big enough to account for a significant change in the norm of the eccentricity, i.e.
\begin{align}
    e^2=e_F^2+e_R^2+2e_Re_F\cos{\theta}. \label{e_norm}
\end{align}

\begin{figure}
   \centering
    \includegraphics[width=0.7\textwidth]{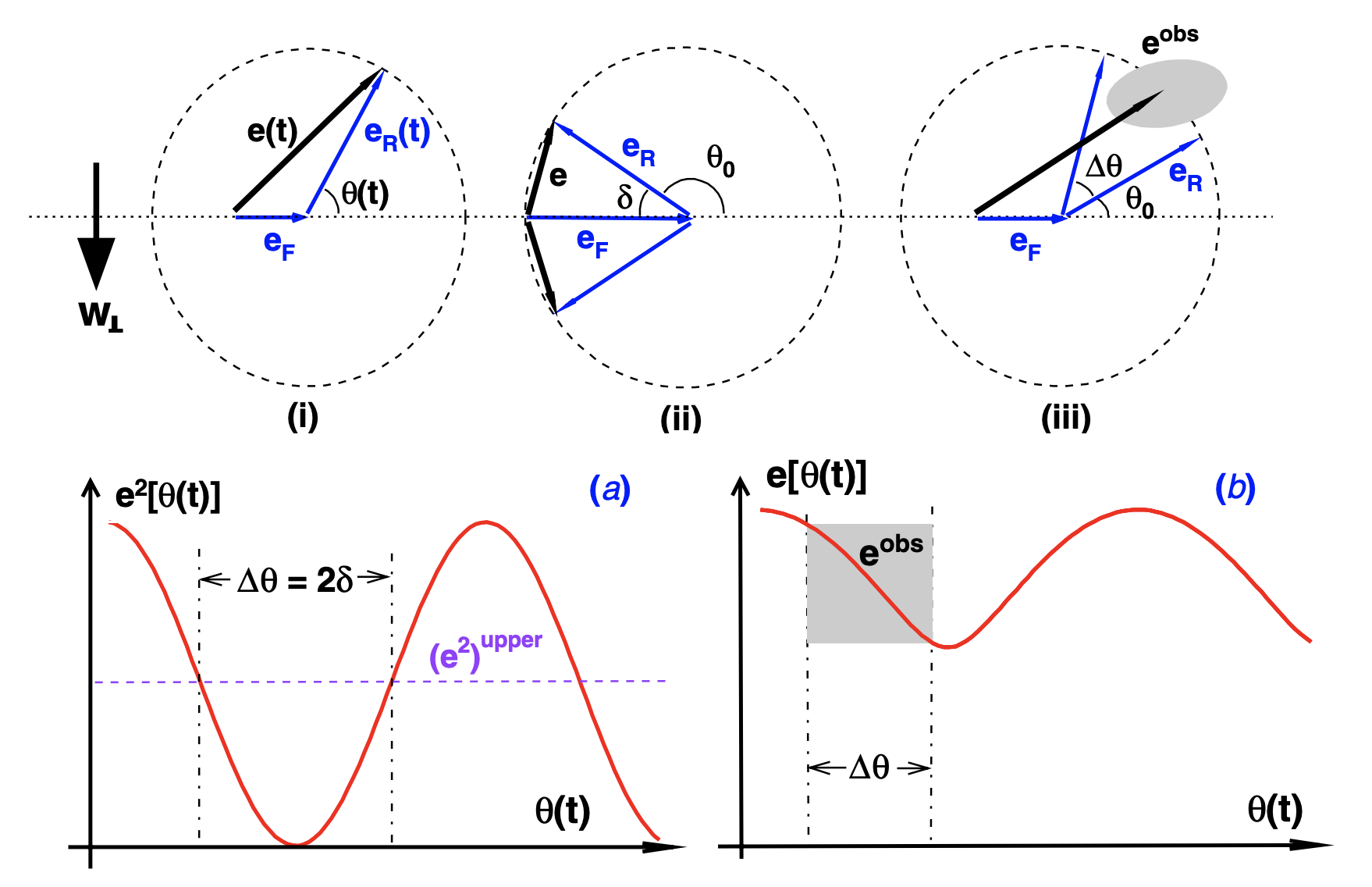}
    \captionof{figure}{Upper : (i) eccentricity vector as a superposition of the precession of periastron component $\vb{e_R}(t)$ and the forced constant component $\vb{e_F}$ at some time $t$, (ii) worst initial condition for the observation of the forced component, the two components cancel out exactly for $\delta=0$, (iii) difference between two observations at time $t_0$ and $t$, $\vb{e_F}$ stays constant whereas $\vb{e_R}$ rotates, $\vb{e_{obs}}$ is the time averaged observed eccentricity vector.\\ Lower : sinusoidal behaviour of the norm of the eccentricity vector with respect to the angle $\theta$. (figure from Shao and Wex 2012 \cite{shao2012new}).\label{alpha1}}
\end{figure}

Note that the norms $e_F$ and $e_R$ are constant and $\theta(t)=\theta_0+\dot{\omega}t$ is linear in time, thus $e^2$ is a sinusoidal function of time. An illustration of the evolution of the eccentricity vector is exposed in figure \ref{alpha1}. One can further note from equation \eqref{e} and \eqref{ef_norm} that the figure of merit to test the PPN parameter $\alphah_1$, i.e. the property required for a good measurements, is \cite{shao2012new} 
\begin{align}
    P_b^2/e.
\end{align}

The observation of eccentricity vector has been conducted for the system PSR J1738+0333 \cite{shao2012new}. This pulsar-white dwarf system is particularly interesting, because its orbital period is short $P_b=8.51$ hours, we know its very low eccentricity vector with a certainty up to $3\sigma$, $e=3.4\cdot 10^{-7}$ \cite{shao2012new}, the time of observation is $T\cong 10$ years and $\Delta\theta\cong 16^\circ$. Using the data for the pulsar system PSR J1738+0333, Shao and Wex were able to set a new limit on $\alphah_1$ at $95\%$CL using \eqref{ef_norm} and \eqref{e_norm} to \cite{shao2012new} :
\begin{align}
    \abs*{\alphah_1}_\textrm{upper}=3.5\cdot 10^{-5}. \label{FOMa1}
\end{align}

\subsection{Eccentricity as a Test of $\vb*{\alphah_3}$}

As mentioned in table \ref{table_PPN}, $\alphah_3$ represents a violation of Lorentz invariance and energy momentum conservation. Because of the latter, most theories of modified gravity predict $\alphah_3=0$ and its test is not used to differentiate different theories. Nevertheless, it is an important indicator of the validity of GR and recent experiments were able to set a very constraining limit on this parameter. In the following we will describe the work of Bell and Damour 1996 \cite{bell1996new} who studied the effect of $\alphah_3$ on the eccentricity of binary systems.

In a binary system, a single body will feel an acceleration due to Newtonian and relativistic effects from the other body in the system. However, for a non-zero $\alphah_3$, each body will feel a  self-acceleration $\vb{a}$ due to violation of energy-momentum conservation \cite{bell1996new}:
\begin{align}
    \vb{a}=-\frac{1}{3}\alphah_3\frac{E_{grav}}{m}(\vb{w}+\vb{v})\times \vb*{\omega},
\end{align}
where $E_{grav}$ is the gravitational self-energy, $m$ the mass, $\vb{w}$ the absolute velocity of the center of mass relative to a preferred frame, $\vb{v}$ is the relative velocity of the object in the system and $\vb*{\omega}$ is the spin angular momentum. In a pulsar-white dwarf system, the perturbation of the relative acceleration $\vb{a_r}=\vb{a_1}-\vb{a_2}$, where the indices $1$ and $2$ represent the pulsar and the white dwarf respectively, can be decomposed as \cite{bell1996new} :
\begin{align}
    \vb{a_r}=\frac{\alphah_3}{6}\vb{w}\times (c_1\vb*{\omega_1}-c_2\vb*{\omega_2})+\frac{\alphah_3}{6}\vb{v}\times (\frac{m_1}{m_1+m_2}c_1\vb*{\omega_1}+\frac{m_2}{m_1+m_2}c_2\vb*{\omega_2}),\label{ar}
\end{align}
where $c_i\cong -2E_{grav,i}/m_i$ is the compactness parameter. For binary pulsars, one usually expect that the spin and orbital angular momentum are parallel, thus $\vb{v}\times\vb*{\omega_i}$ is centripetal and the second term in the RHS of \eqref{ar} gives a non-observable contribution. Furthermore because $c_1\gg c_2$ and $\abs*{\vb*{\omega_1}}\gg\abs*{\vb*{\omega_2}}$ we can reduce the perturbation of the relative acceleration to \cite{bell1996new}
\begin{align}
     \vb{a_r}\cong\frac{\alphah_3}{6}\vb{w}\times c_1\vb*{\omega_1}.
\end{align}
This perturbation will have an influence on the eccentricity vector, $\vb{e}(t)=\vb{e_R}(t)+\vb{e}_{\alphah_3}$, where $\vb{e_R}(t)$ rotates due to periastron precession $\dot{\omega}$, and $\vb{e}_{\alphah_3}$ can be rewritten using $\vb{v}\times\vb*{\omega_1}=w\omega_1\sin\beta\vb{\hat{e}}$ and $\omega_1=2\pi/P$, $P$ the rotation period, as \cite{bell1996new} 
\begin{align}
    \vb{e}(t)=\vb{e_R}(t)+\alphah_3\frac{c_1wP_b^2}{24\pi PM}\sin\beta\vb{\hat{e}}.\label{ee}
\end{align}
Thus by observing $e_{obs}=\abs*{\vb{e}(t)}$ at a certain time $t$ of several low eccentricity systems, one can infer a limit for the value of $\alphah_3$, assuming that $\beta$ is uniformly distributed. One can also see from equation \eqref{ee} that we are looking for a pulsar binary system such that
\begin{align}
    \frac{P_b^2}{eP}\label{FOMa3}
\end{align}
is high, in order to optimize the constraint on the PPN parameter. Using the observation of 9 different binary systems, Bell and Damour 1996 were able to set an upper limit on $\alphah_3$ at $95\%$ CL \cite{bell1996new} :
\begin{align}
    \abs*{\alphah_3}_{upper}=2.4\cdot 10^{-20}.
\end{align}
It is interesting to note that Stairs et al 2005 \cite{stairs2005discovery} followed a similar method using three different newly discovered wide orbit binary pulsars and obtained $\abs*{\alphah_3}_{upper}=4.0\cdot 10^{-20}$, which increases our confidence on this very restraining result.

\subsection{Induced Spin Precession and the limit on $\vb*{\alphah_2}$}\label{induced_spin_precession}

\begin{figure}
   \centering
    \includegraphics[width=0.5\textwidth]{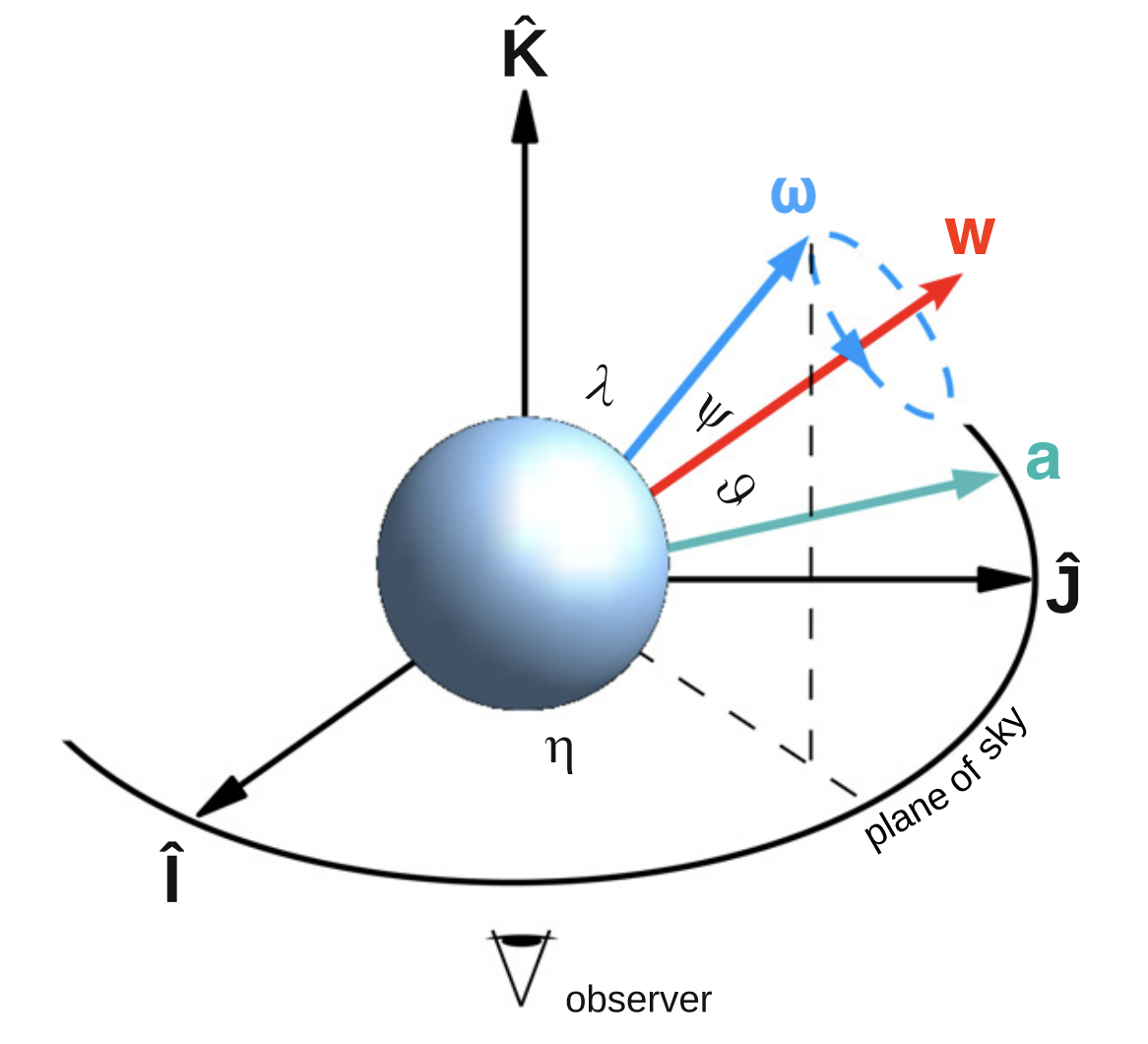}
    \captionof{figure}{ Illustration of a pulsar moving with velocity $\vb{w}$ relative to the preferred frame and spinning with angular velocity $\vb*{\omega}$. The coordinate system $(\vb{\hat{I}},\vb{\hat{J}},\vb{\hat{K}})$ is such that $\vb{\hat{K}}$ is the line of sight and $\vb{\hat{I}}$ and $\vb{\hat{J}}$ are in the plane of sky perpendicular to the line of sight. $\vb{a}=\vb{\hat{K}}\times \vb*{\omega}/\abs*{\vb{\hat{K}}\times \vb*{\omega}}$ lies on the plane of sky.(figure from Shao et al 2013 \cite{shao2013new}).\label{pulsar_spin}}
\end{figure}

Following the reasoning of Nordtvedt 1987 \cite{nordtvedt1987probing} we first describe the effect of $\alphah_2$ on the spin of a pulsar. The relevant contribution to post-Newtonian Lagrangian is of the form
\begin{align}
    L=-\frac{\alphah_2}{2}\sum\limits_{i,j}\frac{m_im_j}{r_{ij}}(\vb{w}\cdot\vb{\hat{r}_{ij}})^2,\label{deltaL}
\end{align}
where $m_i$ are masses of the objects in the system, $\vb{\hat{r}_{ij}}$ is the unit vector of the distance $\vb{r_{ij}}$ between objects $i$ and $j$, $r_{ij}=\abs*{\vb{r_{ij}}}$ and $\vb{w}$ is the velocity of the system relative to the preferred frame. In the case of a single celestial body in internal equilibrium, one can use the tensor virial relation :
\begin{align}
    \frac{1}{2}\sum\limits_{i,j}\frac{m_im_j}{r_{ij}^3}\vb{r_{ij}}\cdot\vb{r_{ij}}^T=\int p(r)\dd V\cdot I_3+\int \rho(r)\vb{v}(\vb{r})\cdot\vb{v}(\vb{r})^T\dd V, \label{virial}
\end{align}
where we neglect the pressure $p(r)$ compare to the density $\rho(r)$ of the object, $I_3$ is the $3\times 3$ identity matrix and $\vb{v}(\vb{r})$ is the rotational velocity field. Furthermore, one can recognize that the second term of the RHS is the rotational kinetic energy tensor $T_{rot}^{ij}$. One can then substitute \eqref{virial} into \eqref{deltaL} to get 
\begin{align}
    -L=\alphah_2T_{rot}(\vb{w}\cdot \vb*{\omegah})^2,
\end{align}
where $\vb*{\omegah}$ is the unit vector of the spin vector $\vb*{\omega}$. One can then identify $-\delta L$ as being an anisotropic potential energy, which translate into a torque $\abs*{\tau}=\abs*{-\partial\delta L/\partial\phi}$ in the motion of the body and thus a spin precession around $\vb{w}$
\begin{align}
    \Omega=2\alphah_2\frac{T_{rot}}{J}w^2\vb{\hat{w}}\cdot\vb*{\omegah}=\alphah_2\omega w^2\cos\psi,  \label{precession}
\end{align}
where $\psi$ is the angle between $\vb{w}$ and $\vb*{\omega}$. Note that in Shao et al 2013 \cite{shao2013new} they use an other parametrization, namely $\left.\alphah_2\right|_{Nordtvedt}=-1/2\left.\alphah_2\right|_{Shao}$. we represent the different components and angles describing the movement of the pulsar in the figure \ref{pulsar_spin}. 

To test for the PPN parameter $\alphah_2$ we remark that the angle $\lambda$ between the line of sight and the spin axis varies as the spin axis precesses around $\vb{w}$ \cite{shao2013new} :
\begin{align}
    \dv{\lambda}{t}=\Omega \cos\theta,
\end{align}
where $\theta$ is the angle between $\vb{w}$ and $\vb{a}$. The precession of the the spin axis induces a precession of the magnetic dipole and thus the angle between the latter and the line of sight, which has for consequence that the intensity of the pulses measured on Earth will vary.

Because geodetic precession, in a binary sytem, can also influence the intensity of the pulses, Shao et al 2013 \cite{shao2013new} worked with the data of two solitary millisecond pulsars. The main requirement for a good measurement of $\alphah_2$ is a short period of rotation $P$ and a long period of observation $T_{obs}$. In the case of PSR B1937+21 and PSR J1744-1134, the time span of the data is $15$ years and their spin period is $P=1.5578$ms and $4.0745$ms, respectively, thus they are good candidates for constraining the PPN parameter. Using both pulsars, Shao et al were able to set a new constraint on the strong field PPN parameter with $95\%$ CL \cite{shao2013new} :
\begin{align}
    \abs*{\alphah_2}_{upper}=1.6\cdot 10^{-9}.
\end{align}

\subsection{Spin Precession Method for the Upper Limit on $\vb*{\xih}$}

As we have seen in the table \ref{table_PPN}, $\xih$ is the strong field analog to the PPN parameter $\xi$ related to the preferred-location effects and it represents the violation of position invariance in the SEP. If we suppose a system composed of a solitary pulsar in the gravitational field of a galaxy, the $\xih$-related term in the Lagrangian is of the form \cite{shao2013xi}
\begin{align}
    L_{\xih}=\frac{\xih}{2}U_G\sum\limits_{i,j}\frac{m_im_j}{r_{ij}^3}(\vb{r_{ij}}\cdot \vu{n}_G)^2,
\end{align}
where $U_G$ is the gravitational potential of the galaxy and $\vu{n}_G=\vb{R}_G/R_G$ is the unit vector between the center of the galaxy and the solitary pulsar. We see a similar structure as in \eqref{deltaL} and following the development of Nordtvedt 1987 \cite{nordtvedt1987probing}, one can show that it will translate into a spin precession as in the case of $\alpha_2$ \cite{shao2013xi} :
\begin{align}
    \Omega=\xih\frac{2\pi}{P}v_G\cos\psi,
\end{align}
where $P$ is the spin period, $v_G=\sqrt{U_G}$ is an effective velocity and $\psi$ is the angle between spin axis and $\vu{n}_G$. Note here again that the parametrization between Shao and Wex \cite{shao2013xi} and Nordtvedt \cite{nordtvedt1987probing} is different : $\left.\xih\right|_{Nordtvedt}=-1/2\left.\xih\right|_{Shao}$.

Following a similar method as exposed in section \ref{induced_spin_precession}, Shao and Wex 2013 \cite{shao2013xi} found an upper limit for the strong field generalization $\xih$ at $95\%$ CL :
\begin{align}
    \abs*{\xih}_{upper}=3.9\cdot 10^{-9}.
\end{align}

\subsection{Nordtvedt Effect and the Universality of Free Fall}\label{section_nordtvedt}
We have yet not talked about the weak equivalence principle, which states that all objects experience the same acceleration in a given gravitational field, in other word, the ratio between the inertial mass $m_i$ and gravitational mass $m_g$ is constant. This ratio is usually normalized to one using Newton's constant. We can express this ratio in terms of the PPN parameters as \cite{stairs2003testing}
\begin{align}
    \frac{m_g}{m_i}:=1+\Delta=1+\eta\left(\frac{E_g}{m}\right)+\eta'\left(\frac{E_g}{m}\right)^2+...,
\end{align}
where $E_g$ is the gravitational self-energy, $m$ is the total mass energy and 
\begin{align}
    \eta=4\beta-\gamma-3-\frac{10}{3}\xi-\alpha_1+\frac{2}{3}\alpha_2-\frac{2}{3}\zeta_1-\frac{1}{3}\zeta_2
\end{align}
is the Nordtvedt parameter. At first order we thus have $\Delta=\eta E_g/m$ and a non-zero Nordtvedt parameter would lead to different acceleration for bodies with different fractional gravitational self-energy $F:=E_g/m$. Extensive tests have been conducted for the Earth-Moon system in the gravitational field of the Sun, however, because the difference between $F_{earth}$ and $F_{moon}$ is of order $10^{-10}$, even with a  precise measurement of the relative acceleration $\abs{\Delta}=\abs{\Delta_{earth}-\Delta_{moon}}$ due to the Sun gravitational field, it is hard to constraint $\eta$. Measurements using the Lunar Laser Ranging lead to a remarkable constraint of $\abs{\eta}<2.9\cdot 10^{-4}$ \cite{hofmann2018relativistic}. 

One can also try to constraint the Nordtvedt parameter using a pulsar and its companion in the gravitational field of their galaxy, however the recent discovery of the triple system PSR J0337+1715, allows for a more precise measurement of the parameter. Here again, since we are dealing with compact objects, we need to consider the strong field analog $\etah$ of the Nordtvedt parameter and the comparison with the LLR result is not straight forward. The interesting particularity of the triple system is that the pulsar has two white dwarf companions, the companion $A$ with an orbital period of $P_b=1.63$ days and the comanion $B$ orbiting further with $P_b=327.26$ days. Archibald et al 2018 \cite{archibald2018universality} were able to constraint the relative acceleration between the pulsar and the companion $A$ in the gravitational field of the companion $B$, $\abs{\Delta_{P}-\Delta_A}<2.6\cdot 10^{-6}$. Since the difference $F_P-F_A\sim 0.1$ it leads to a constraint at $95\%$ CL of the strong field Nordtvedt parameter \cite{archibald2018universality}
\begin{align}
    \abs{\etah}_{upper}=2.6\cdot 10^{-5}.
\end{align}

\subsection{Variation of Newton's Constant}

Some modified gravity theories predict a variation of the Newton's constant $G$ with respect to a preferred cosmic frame. In order to correctly describe the variation of $G$, we don't normalize it to $1$ in this subsection. The variation is usually of the form \cite{will2018theory}
\begin{align}
    \frac{\gdot}{G}\cong \sigma H_0,
\end{align}
where $H_0$ is the Hubble constant and $\sigma$ is a dimensionless parameter depending on the gravity theory. In particular, we have $\sigma=0$ for General Relativity. Such a variation should have an impact on the orbital period of a binary system and on the spin of a compact object. In this section, we will analyze in detail the $\gdot$ effect on the orbital period of a binary system. We can decompose the evolution of the orbital period $\dot{P}_b$ into five contributions \cite{lazaridis2009generic}
\begin{align}
    \pdot_b=\pdot_{\dot{m}} +\pdot_T+\pdot_D+\pdot_{GW}+\pdot_{\gdot},\label{ppppp}
\end{align}
where $\pdot_{\dot{m}}$ comes from the loss of mass due to loss of rotational energy, $\pdot_T$ is the tidal force losses contribution, $\pdot_D$ takes into account the Doppler correction, $\pdot_{GW}$ is the loss of energy through gravitational waves and finally $\pdot_{\gdot}$ is the $\gdot$ effect on the orbital motion. The two first terms are neglected because they are small compared to $\pdot_b$. The value for $\pdot_b$, $\pdot_D$ and $\pdot_{GW}$ are calculated for the binary pulsar PSR J1012+5307 in Lazaridis et al 2009 \cite{lazaridis2009generic}. In order to constraint $\gdot/G$ we need to describe the contribution $\pdot_{\gdot}$, which we derive following Nordvedt 1990 \cite{nordtvedt1990g}. We first write the orbital period of a pulsar at post-Newtonian order for a general gravity theory :
\begin{align}
    P_{\gdot}\cong\frac{2\pi l^3}{G^2m^2}(1-e^2)^{-3/2},\label{period}
\end{align}
where $l=r^2\dot{\theta}$ is the usual angular momentum, $e$ the eccentricity and $m=m_1+m_2$, with $m_1$, $m_2$ the mass of the pulsar and the companion, respectively. Using \eqref{period} one can derive the orbital period evolution due to a variation of Newton's constant
\begin{align}
    \frac{\pdot_{\gdot}}{P_{\gdot}}=3\frac{\dot{l}}{l}-2\frac{\dot{G}}{G}-2\frac{\dot{m}}{m}.\label{pdotp}
\end{align}
The mass term is typically a function of the Newton's constant and the PPN parameters, it can be written at first order as
\begin{align}
    \frac{\dot{m}}{m}=\frac{G}{m}\frac{\delta m}{\delta G}\frac{\gdot}{G},\label{mdotm}
\end{align}
where
\begin{align}
    \frac{\delta m}{\delta G}=\pdv{m}{G}+\pdv{m}{\gamma}\frac{\dot{\gamma}}{\gdot}+\pdv{m}{\beta}\frac{\dot{\beta}}{\gdot}+...
\end{align}
We further define the compactness $c_i$ of an object $i=1,2$ as 
\begin{align}
    c_i:=\frac{G}{m_i}\frac{\delta m_i}{\delta G},
\end{align}
thus equation \eqref{mdotm} becomes
\begin{align}
    \frac{\dot{m}}{m}=\frac{m_1c_1+m_2c_2}{m_1+m_2}\frac{\gdot}{G}.\label{mdotm2}
\end{align}
We now want to describe the angular momentum term in \eqref{pdotp} using $\gdot$ and $G$. If we only consider the effect of $\gdot$ on the orbital period, which is our case, then we can use the conservation of momentum $\dot{\vb{p}}=0$ to write 
\begin{align}
    \dv{\vb{v}_i}{t}=-\frac{\dot{m}_i}{m_i}\vb{v}_i=-\frac{\gdot}{G}c_i\vb{v}_i,\label{accel}
\end{align}
where we further assume non-relativistic speed and $\vb{v}_i$ is the speed of an object $i$ relative to the cosmos. Note that this acceleration is only caused by the variation of $G$ if the body is moving in the cosmos. We can decompose the velocity in two parts $\vb{v}_i=\vb{u}_i+\vb{w}$, where $\vb{u}_i$ is the velocity of object $i$ relative to the center of mass and $\vb{w}$ is the velocity of the center of mass relative to the cosmos. Furthermore, equation \eqref{accel} will induce a change in angular momentum $\vb{l}$ of the binary system. If $\vb{l}=\vb{r}\times \vb{v}$, where $\vb{r}=\vb{r}_1-\vb{r}_2$ is the distance between the two objects and $\vb{v}=\dot{\vb{r}}$, then
\begin{align}
    \dv{\vb{l}}{t}&=\vb{r}\times \vb{a}=-\frac{\gdot}{G}\vb{r}\times(c_1\vb{v}_1-c_2\vb{v}_2)\nonumber\\
    &=-\frac{\gdot}{G}((c_1-c_2)\vb{r}\times\vb{w}+c_1\vb{l}_1+c_2\vb{l}_2)\nonumber\\
    &=-\frac{\gdot}{G}\left((c_1-c_2)\vb{r}\times\vb{w}+\frac{c_1m_2+c_2m_1}{m_1+m_2}\vb{l}\right),\label{dl}
\end{align}
where $\vb{l}_i=\vb{r}_i\times\vb{u}_i$ is the angular momentum of object $i$ with respect to the center of mass. Lazaridis et al 2009 \cite{lazaridis2009generic} considered the pulsar PSR J1012+5307, which has a white dwarf companion. This system has a very small eccentricity, $e<8.4\cdot 10^{-7}$, we can assume that the orbits are circular and $\vb{r}\times \vb{w}=\vb{0}$. Furthermore the compactness $c_2$ of the white dwarf is negligible compared to the compactness $c_1$ of the pulsar, thus, using equation \eqref{dl}, one can write 
\begin{align}
    \frac{\dot{l}}{l}=-\frac{c_1m_2}{m_1+m_2}\frac{\gdot}{G}.\label{ldotl}
\end{align}
Finally we can substitute \eqref{mdotm2} and \eqref{ldotl} into \eqref{pdotp} to obtain
\begin{align}
    \frac{\pdot_{\gdot}}{P_{\gdot}}=-2\frac{\gdot}{G}\left(1+(1+\frac{m_1}{2m})c_1\right).
\end{align}
Note that in Lazaridis et al 2009 \cite{lazaridis2009generic} they use the notion of sensitivity $s_1$ instead of compactness, both notions are connected as $s_1=-c_1$. Finally using equations \eqref{ppppp} and \eqref{ldotl}, together with the measured or calculated values of $P_b$, $P_D$, $P_{GW}$, $m_1$, $m_2$ and $c_1$ for PSR J1012+5307, one can find a bound for the variation of Newton's constant. Lazaridis et al 2009 \cite{lazaridis2009generic} could measure a value for $\gdot/G$ at $95\%$ CL of 
\begin{align}
    \frac{\gdot}{G}=(-0.7\pm 3.3)10^{-12}.
\end{align}
This result seems very constraining for the variation of Newton's constant and indeed it is one of the lowest measured value for $\gdot/G$. However measurements conducted in the Solar system could set a stronger restriction on this variation. In particular, Konopliv et al 2011 \cite{konopliv2011mars} and  Pitjeva and Pitjev 2013 \cite{pitjeva2013relativistic} found a value of $\gdot/G=(0.1\pm 1.6)10^{-13}$ and $\gdot/G=(-0.6\pm 0.4)10^{-13}$ (values taken from Will 2018 \cite{will2018theory}), respectively. Note that the latter found a significant deviation from zero, however most of the tests include the null value in their confidence interval and one could expect a false test at $95\%$ CL, if $\gdot/G$ is indeed zero. Further more precise measurements should be conducted to investigate this deviation.

\subsection{First Discussion on SEP Tests}
We start by summarizing the current lowest upper limit of the PPN parameters the Nordtvedt parameter and the variation of the Newton's constant representing violations of the SEP in the table \ref{constaints}. As already mentioned the limits were set on strong field analog to the PPN parameters, however one can infer a limit on the weak field parameters using the reasonable assumption that strong field corrections would increase the deviation on the PPN parameters. For more information on the weak field tests, see Will 2014 \cite{will2014confrontation}.

\begin{center}
\begin{tabular}{|c|c|c|}
\hline
  Parameters          & Value               & Method                              \\ \hline\hline
$\abs*{\alphah_1}$ & $3.5\cdot 10^{-5}$  & Eccentricity of binary pulsars      \\ \hline
$\abs*{\alphah_2}$ & $1.6\cdot 10^{-9}$  & Eccentricity of binary pulsars      \\ \hline
$\abs*{\alphah_3}$  & $2.4\cdot 10^{-20}$ & Spin precession of solitary pulsars \\ \hline
$\abs*{\xih}$    & $3.9\cdot 10^{-9}$  & Spin precession of solitary pulsars \\ \hline
$\abs*{\etah}$    & $2.6\cdot 10^{-5}$  & Nordtvedt effect \\ \hline
\multirow{2}{*}{$\abs*{\gdot/G}$} & $4\cdot 10^{-12}$ & Variation of orbital period of binary pulsar \\
                           & $1.7 \cdot 10^{-13}$ & Mars ephemeris                             \\ \hline
\end{tabular}
\captionof{table}{Summary of the lowest upper limit with $95\%$ CL of the PPN parameters, the Nordtvedt parameters and the variation of Newton's constant representing violations of the strong equivalence principle and the method used to set the limit. The hat on the PPN parameters means that we are considering the strong field analog to the weak field parameters described in section \ref{section_PPNF}. \label{constaints}}
\end{center}

The main conclusion one can make using the tests we described in this section is that no significant deviation from general relativity has been discovered for the strong equivalence principle using the parametrized post-Newtonian formalism and the Keplerian parameters. In this review, we only considered the most stringent tests, however there were a lot of studies testing the parameters summarized in table \ref{constaints} using similar methods and none of them reported a significant deviation from zero (see conclusion of Will 2014 \cite{will2014confrontation}). The result of Pitjeva and Pitjev 2013 \cite{pitjeva2013relativistic} of the variation of Newton's constant is significantly different than zero, $\gdot/G=(-0.6\pm 0.4)\cdot10^{-13}$ at $95\%$ CL. If $G$ is indeed a constant, one might still expect some false significant deviations from zero for $\gdot$ at $95\%$ CL. However further tests with higher precision should be conducted in order to falsify or verify this result. 

If one want to set stronger constraints on the parameters described in table \ref{constaints}, one can start by finding better candidates for the tests being considered. In the case of $\alphah_1$ and $\alphah_3$, binary pulsars with a high figure of merit described by \eqref{FOMa1} and \eqref{FOMa3}, respectively, should be studied. For $\alphah_2$ and $\xih$, a solitary pulsar with very high rotation frequency and long period of observation is required for a good test. Better candidates than the triple system PSR J0337+1715 to constraint the Nordtvedt parameter might be hard to find, one can look for an other triple system or binary systems close to the center of their galaxy. Finally for the variation of Newton's constant, one need a binary system with a very low eccentricity, short orbital period and long observation time to have a precise measurement of $\dot{P}_b$.

%--------------------------

\section{Scalar-Tensor Theories} \label{section_ST}
Scalar-tensor theories (ST theories) are among the best alternative candidates of General Relativity. The first ST theory was initiated by Brans and Dicke in 1961 and a lot of different versions have florished at the end of the 20th century, following the trend of string theory, with the dilaton field, or inflationary models. The idea behind ST theories is simple, adding a scalar field $\phi$ together with the usual tensor field $g_\munu$. The action of scalar-tensor theories can be written as \cite{will2018theory}
\begin{align}
    S[\phi,g_\munu]=\frac{1}{16\pi G}\int \left(\phi R-\frac{\omega (\phi)}{\phi}g^\munu \phi_{,\mu}\phi_{,\nu}-U(\phi)\right)\sqrt{-g}\dd^4x+S_M[\psi_m,\phi,g_\munu], \label{action1}
\end{align}
where $\omega(\phi)$ is an arbitrary coupling function, $R$ is the Ricci scalar, $U(\phi)$ is an arbitrary potential and $S_M$ is the action of matter coupling. There are several manipulations one can perform to make \eqref{action1} look nicer. We start by redefining the tensor field $g_\munu\rightarrow \tilde{g}_\munu=\phi/\phi_0 g_\munu$, where $\phi_0$ is a normalization constant, supposedly the value of the scalar field far away. This gives a new action
\begin{align}
   S[\phi,\gt_\munu]=\frac{1}{16\pi \tilde{G}}\int \left(\tilde{R}-\frac{3+2\omega (\phi)}{2\phi^2}\gt_\munu \phi_{,\mu}\phi_{,\nu}-V(\phi)\right)\sqrt{-\gt}\dd^4x+S_M[\psi_m,\phi,\phi^{-1}\gt_\munu],\label{action2}
\end{align}
where $\tilde{R}$ is the Ricci scalar associated to $\gt_\munu$, $V(\phi)=\phi_0U(\phi)/\phi^2$ and $\tilde{G}=G/\phi_0$. Furthermore we can also redefine the scalar field, $\phi\rightarrow\varphi$, such that 
\begin{align}
    A(\varphi)&=\phi^{-1/2},\label{eq_defA}\\
    \dv{A(\varphi)}{\varphi}&=\phi^{-1/2}(3+2\omega(\phi))^{-1/2}\label{eq_dA}
\end{align}
and the action becomes \cite{will2018theory}
\begin{align}
    S[\varphi,\gt_\munu]=\frac{1}{16\pi \tilde{G}}\int\left(\tilde{R}-2\gt^\munu\varphi_{,\mu}\varphi_{,\nu}-\bar{V}(\varphi)\right)\sqrt{-\gt}\dd^4 x+S_M[\psi_m,\varphi,A^{2}(\varphi)\gt_\munu],\label{action3}
\end{align}
where $\bar{V}(\varphi)=V(\phi)$. In the following sections, we will construct the PPN and PPK formalism for the theory described by the action \eqref{action2} and expose some interesting tests of this theory using pulsar timing.

\subsection{PPN Formalism in ST Theories}\label{sect_PPNST}
We can now use the PPN formalism described in section \ref{section_PPN} in the case of scalar-tensor theories and compare the result with General Relativity. For this purpose, we summarize the development of Will 2018 \cite{will2018theory} and for more details see its chapter 5. From \eqref{action2} we can write the field equations for $\gt_\munu$ and $\phi$, using $\tilde{h}^\munu:=\eta^\munu-\sqrt{-\gt}\gt^\munu$, with $\eta_\munu=\textrm{diag}\{-1,1,1,1\}$, and imposing the harmonic coordinates condition $\tilde{h}^\munu_{,\nu}=0$
\begin{subequations}
\begin{align}
    \square \tilde{h}^\munu&=-16\pi\tilde{G}(-\gt)\left(\tilde{T}^\munu+\tilde{t}^\munu_\phi+\tilde{t}^\munu_{LL}+\tilde{t}^\munu_H\right),\label{eqeomh}\\
    \square \phi&=-8\pi \tilde{G}\tau_s,\label{eqeomphi}
\end{align}
\end{subequations}
where $\tilde{T}^\munu=(\phi_0/\phi)^3T^\munu$, $\tilde{t}^\munu_{\phi}$, $\tilde{t}^\munu_{LL}$ and $\tilde{t}^\munu_{H}$ are the field, the Landau-Lifshitz and the harmonic pseudo-tensors, respectively, they are described in Will 2018 \cite{will2018theory} as well as $\tau_s$. We further assume perfect fluid thus $T_\munu=(\rho+\rho\Pi+p)u_\mu u_\nu+g_\munu p$. To solve equations \eqref{eqeomh} and \eqref{eqeomphi}, we can use the iterative method, i.e. one substitute the zeroth order $h_0^\munu=0$ and $\phi_0$ in the RHS and then solve the equations for the first order $h_1^\munu$ and $\phi_1:=\phi_0\Psi$. The solution at first, Newtonian order is 
\begin{subequations}
\begin{align}
    \tilde{h}_1^{00}&=4\tilde{G}U \\
    \tilde{h}_1^{0j}&=4\tilde{G}V^j \\
    \tilde{h}_1^{ik}&= 0, \quad i\neq k\\
    \Psi&= \frac{2}{3+2\omega_0}\tilde{G}U,
\end{align}
\end{subequations}
where $\omega_0=\omega(\phi_0)$ and 
\begin{align}
    V^j=\int \rho(t,\vb{x}')\frac{v'^j}{\abs*{\vb{x}-\vb{x}'}}\dd^3x'.
\end{align}
using the Newtonian approximation $g_{00}=-1+2GU$, we can describe the modified Newton's constant $\tilde{G}$ using the usual Newton's constant $G$ as 
\begin{align}
    \tilde{G}=G\frac{3+2\omega_0}{4+2\omega_0}.
\end{align}
One can now repeat the process to compute the second, post-Newtonian order and the post-Newtonian metric for scalar-tensor theories can be written as 
\begin{subequations}
\label{eq_STmetric}
\begin{align}
    g_{00}&=-1+2U+2\left(\psi-(1+\zeta \lambda)U^2+\half \ddot{X}\right)+O(\epsilon^3)\\
    g_{0j}&=-\frac{4}{c^3}(1-\zeta)V_j+O(\epsilon^{5/2})\\
    g_{ik}&=\delta_{ik}(1+2(1-2\zeta)U)+O(\epsilon^2),\quad i\neq k\\
    \Psi&=2\zeta U+O(\epsilon^2),
\end{align}
\end{subequations}
where 
\begin{align}
    \psi=\half(3-4\zeta)\Phi_1-(1+2\zeta\lambda)&\Phi_2+\Phi_3+3(1-2\zeta)\phi_4,\nonumber\\
    \zeta=\frac{1}{4+2\omega_0},\qquad \lambda=\frac{\phi_0\omega'(\phi_0)}{(3+2\omega_0)(4+2\omega_0)}&,\quad X(t,\vb{x})=\int \rho(t,\vb{x}')\abs*{\vb{x}-\vb{x}'}\dd^3x' \nonumber\\
    \grad^2\Phi_1=-4\pi\rho v^2,\quad \grad^2\Phi_2=-4\pi\rho U,\quad &\grad^2\Phi_3=-4\pi\rho \Pi,\quad \grad^2\Phi_3=-4\pi p.\nonumber
\end{align}
We find the PPN parameters for scalar-tensor theories by comparing the general PPN metric \eqref{eq_PPNmetric} with the post-Newtonian metric for ST theories \eqref{eq_STmetric} in the standard PPN gauge, i.e. $t=\bar{t}+\dot{X}/2$ and $x^j=\bar{x}^j$. The result gives 
\begin{align}
    \gamma=\frac{1+\omega_0}{2+\omega_0},\quad \beta=1+\frac{\phi_0\omega'(\phi_0)}{(3+2\omega_0)(4+2\omega_0)^2}\label{eq_PPNST}
\end{align}
and $\xi=\zeta_i=\alpha_i=0$. From here, we see that most of the tests we have seen in the section \ref{section_SEP} are not constraining scalar-tensor theories, except the variation of Newton's constant and the Nordtvedt effect. 

\subsection{Coupling Function and the 2D Space of Tensor-Scalar Theories}
\begin{figure}
   \centering
    \includegraphics[width=0.65\textwidth]{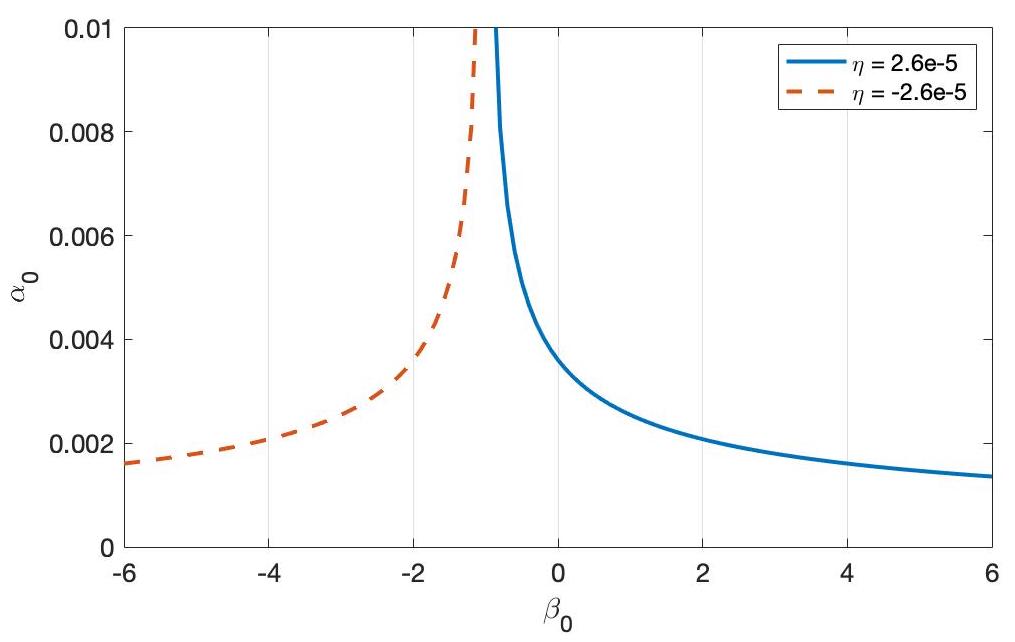}
    \captionof{figure}{Constraint on the asymptotic coupling constant $\alpha_0$ and $\beta_0$ of scalar-tensor theories using the triple system PSR J0337+1715 and the constraint on the strong field Nordtvedt parameter $\etah$ exposed in table \ref{constaints}. The dashed line represents the upper bound of $\alpha_0(\beta_0)$ for the limit $\eta(\alpha_0,\beta_0)\geq -\abs*{\etah}_{upper}$ and the solid line is the upper limit for  $\eta(a_0,b_0)\leq \abs*{\etah}_{upper}$, where $\abs*{\etah}_{upper}=2.6\cdot 10^{-5}$ is the constraint found by Archibald et al 2018 \cite{archibald2018universality} on the strong field Nordtvedt parameter. \label{fig_a0b0constraint}}
\end{figure}

Our purpose is now to describe all the space of all the different scalar-tensor theories derived from the Lagrangian \eqref{action3} at post-Newtonian order. The space we are about to construct also includes more general Lagrangian as explained in Damour 2009 \cite{damour2009binary}. We start by defining the coupling function $a(\varphi):=\log(A(\varphi))$. The name of this function has not been randomly chosen, we can describe the coupling of the matter fields through gravity due to the exchange of a scalar field, which gives us a deviation from GR. Later on we want to expand the coupling function up to second order, thus we further define $\alpha(\varphi):=a'(\varphi)$ and $\beta(\varphi):=a''(\varphi)$. Using the definition of $a(\varphi)$, we can derive the expressions for $\alpha(\varphi)$ and $\beta(\varphi)$. Starting with the former we have
\begin{align}
    \alpha(\varphi)=\frac{A'(\phi)}{A(\phi)}=(2\omega(\phi)+3)^{-1/2},\label{eq_couplinga}
\end{align}
using the equations \eqref{eq_defA} and \eqref{eq_dA}. To find the expression for $\beta(\varphi)$ we first need 
\begin{align}
    \pdv{\phi}{\varphi}=\pdv{(A^{-2}(\varphi))}{\varphi}=-2\alpha(\varphi)\phi.\label{eq_defa}
\end{align}
The expression of $\beta(\varphi)$ is given by
\begin{align}
    \beta(\varphi)=\alpha'(\varphi)=-\frac{\omega'(\phi)}{(3+2\omega(\phi))^{3/2}}\pdv{\phi}{\varphi} =\frac{2\omega'(\phi)\phi}{(3+\omega(\phi))^2}.\label{eq_defb}
\end{align}
Now we can take the asymptotic limit $\phi\rightarrow\phi_0$ of the derivatives of the coupling function, namely $\alpha(\varphi_0):=\alpha_0$ and $\beta(\varphi_0):=\beta_0$, and describe the PPN parameters \eqref{eq_PPNST} using only these two variables
\begin{align}
    \gamma=1-2\frac{\alpha^2_0}{1+\alpha_0^2}\quad\textrm{and}\quad \beta=1+\half\frac{\alpha_0^2\beta_0}{(1+\alpha_0^2)^2}.\label{eq_PPNa0b0}
\end{align}
Furthermore we can also write $G=\tilde{G}(1+\alpha_0^2)$. This is quite remarkable, we are able to describe the effect of the scalar field on the post-Newtonian metric using only the two first derivatives of the coupling function. One can interpret this result using the notion of Feynman diagrams. Indeed the interaction between two worldlines is mediated by a graviton and a scalar field. The scalar field couples to matter with strength $\tilde{G}^{1/2}\alpha_0$, thus the exchange of a scalar field between two matter fields gives a contribution $\tilde{G}\alpha_0^2$, due to two vertices. We can thus describe scalar-tensor theories at post-Newtonian order with the second order expansion of the coupling function
\begin{align}
    a(\varphi)=\alpha_0(\varphi-\varphi_0)+\half \beta_0(\varphi-\varphi_0)^2.
\end{align}
We can now define the 2D space of scalar-tensor theories $T(\alpha_0,\beta_0)$, which are described with the coupling function $a(\alpha_0,\beta_0)$. It is important to note that in the weak field limit, this 2D space describes all the possible monoscalar-tensor theories, thus constraining the asymptotic coupling constants amounts to reduce the size of possible theories in $T(\alpha_0,\beta_0)$. Using the constraint on the strong field Nordtvedt parameter $\etah$ exposed in table \ref{constaints}, we can deduce a first constraint on the asymptotic coupling constants. Indeed the Nordtvedt parameter for scalar-tensor theories is 
\begin{align}
    \eta(\alpha_0,\beta_0)=2\alpha_0^2\frac{1+\beta_0+\beta_0^2}{(1+\alpha_0^2)^2}.\label{eq_etaST}
\end{align}
Since we are constraining two parameters with only one equation, the resulting upper bound will be a curve $\alpha_0(\beta_0)$. We can easily see in equation \eqref{eq_etaST} that $\eta(\alpha_0,\beta_0)$ is symmetric in $\alpha_0$, as well as equations \eqref{eq_PPNa0b0}, thus we can only consider the positive values for $\alpha_0$ without loss of generality. The most constraining upper bound for the strong field Nordtvedt parameter  was set by Archibald et al 2018 \cite{archibald2018universality} with the triple system PSR J0337+1715 as described in section \ref{section_nordtvedt}. Using their constraint, we draw the upper limit curve for the asymptotic coupling constants in figure \ref{fig_a0b0constraint}. 
As we could expect from equation \eqref{eq_etaST}, with a small value for $\alpha_0$, $\beta_0$ is very weakly bounded, however it has little effect on the PPN parameter $\beta$ as we can observe in equation \eqref{eq_PPNa0b0}. We can also observe on the figure that for $\beta_0\cong -1$, the maximum value for $\alpha_0$ seems to diverge, however it is not the case. Indeed the equation $\eta(\alpha_0,\beta_0)=-\abs*{\etah}_{upper}$, represented by the dashed line, does not admit solutions for $\beta_0 > -1.01$, where $\alpha_0\cong 0.06$ is the value at the boundary, but with this value for $\beta_0$ the equation $\eta(\alpha_0,\beta_0)=\abs*{\etah}_{upper}$, represented by the solid line, leads to $\alpha_0\cong 0.09$, which is the highest value possible for $\alpha_0$. Once more we need to be careful when analyzing this result, $\eta(\alpha_0,\beta_0)$ was derived using the weak field limit, whereas $\abs*{\etah}_{upper}$ was set using a triple system in the strong field regime, thus other parameters than $\alpha_0$ and $\beta_0$ might be needed to completely describe the system using scalar-tensor theories. 

\subsection{Spontaneous Scalarization and the Lower Bound on $\vb*{\beta_0}$}
We start by describing the internal problem of a compact object using scalar-tensor theories. For this purpose we will rely on the work of Damour 2009 \cite{damour2009binary}. Using the action \eqref{action3} we can derive the field equations 
\begin{subequations}
\begin{align}
    \tilde{R}_\munu&=2\partial_\mu\varphi\partial_\nu \varphi +8\pi \tilde{G}(\tilde{T}_\munu-\half \tilde{g}_\munu T),\\
    \Box_{\tilde{g}}\varphi&=-4\pi \tilde{G}a(\varphi)\tilde{T},
\end{align}
\end{subequations}
where $\tilde{T}^\munu=2(\tilde{g})^{-1/2}\delta S_M/\delta \tilde{g}_\munu$. One can work on the field equations for a slowly rotating neutron star with the boundary conditions $\tilde{g}_\munu\rightarrow \eta_\munu$ and $\varphi\rightarrow \varphi_0$ at large distances to find the relevant contribution of the scalar field to the total mass $m_A(\varphi_0)$, the total scalar charge $\omega_A(\varphi_0)$ and the moment of inertia $I_A(\varphi_0)$ of the neutron star. Furthermore they satisfy the relation
\begin{align}
    \pdv{m_A}{\varphi_0}=-\frac{\omega_A}{m_A}:=\alpha_A(\varphi_0),
\end{align}
and we define useful quantities
\begin{align}
    \beta_A(\varphi_0):=\pdv{\alpha_A}{\varphi_0}\quad\textrm{and}\quad k_A(\varphi_0):=-\pdv{I_A}{\varphi_0}.
\end{align}
Note that in the case of a weak self-graviting object, the parameter $\alpha_A(\varphi_0)$ becomes the first derivative of the coupling function $\alpha(\varphi_0)$ defined in \eqref{eq_couplinga} and thus shares a similar physical interpretation with $\alpha_0$, being the effective coupling strength between the neutron star and the scalar field. 

\begin{figure}[t]
\begin{minipage}[b]{0.46\textwidth}
\captionsetup{width=1\linewidth}
\centering
\includegraphics[width=0.85\textwidth]{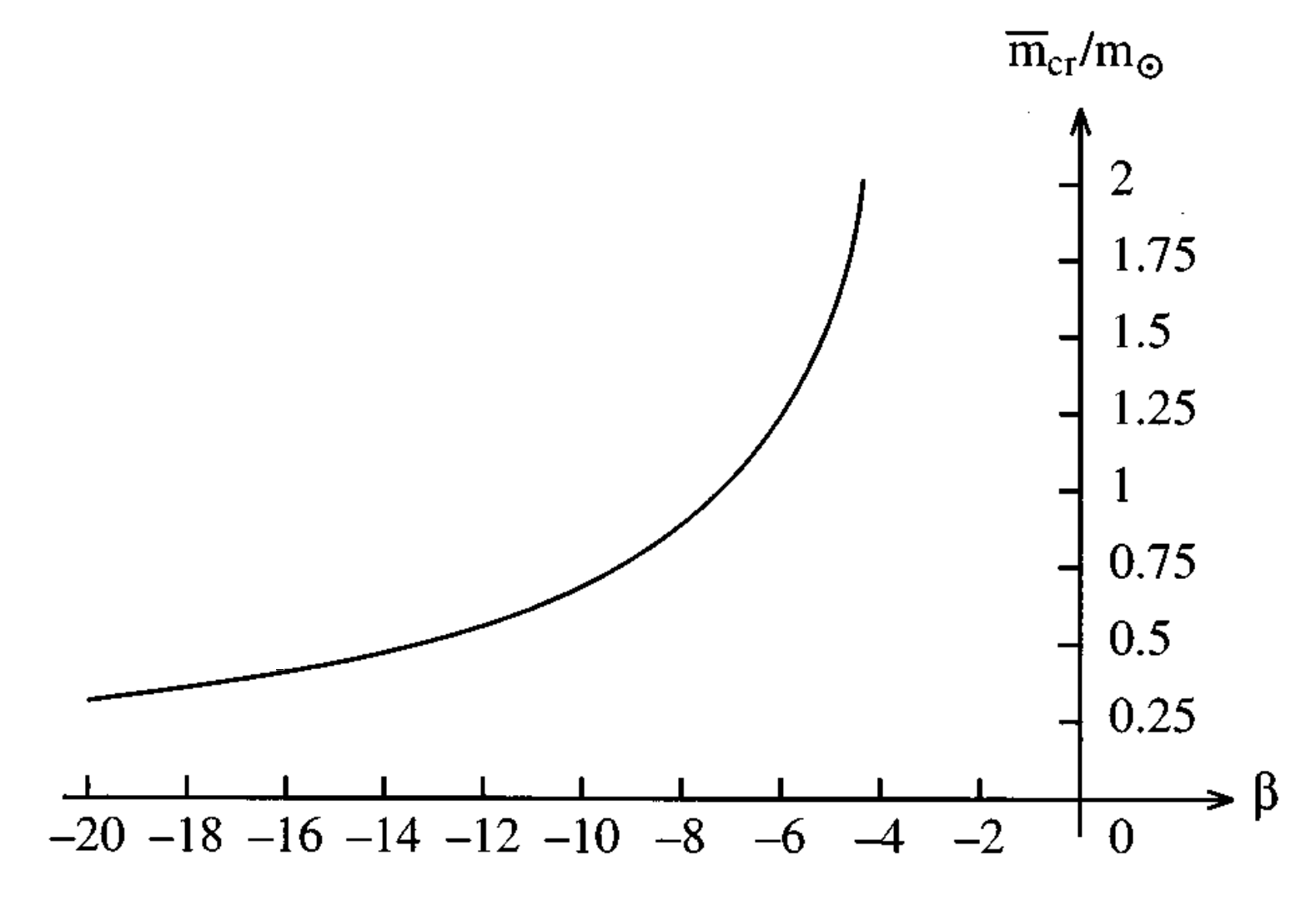}
\captionof{figure}{Dependence of the critical mass to solar mass ratio on the parameter $\beta_0$ for the specific scalar-tensor theory with $A(\varphi)=\exp(\half \beta_0 \varphi^2)$. Figure taken from Damour and Esposito-Farèse 1996 \cite{damour1996tensor}.\newline \newline \label{fig_beta_mass}}
\end{minipage}
\hfill
\begin{minipage}[b]{0.46\textwidth}
\centering
\captionsetup{width=1\linewidth}
\includegraphics[width=1\textwidth]{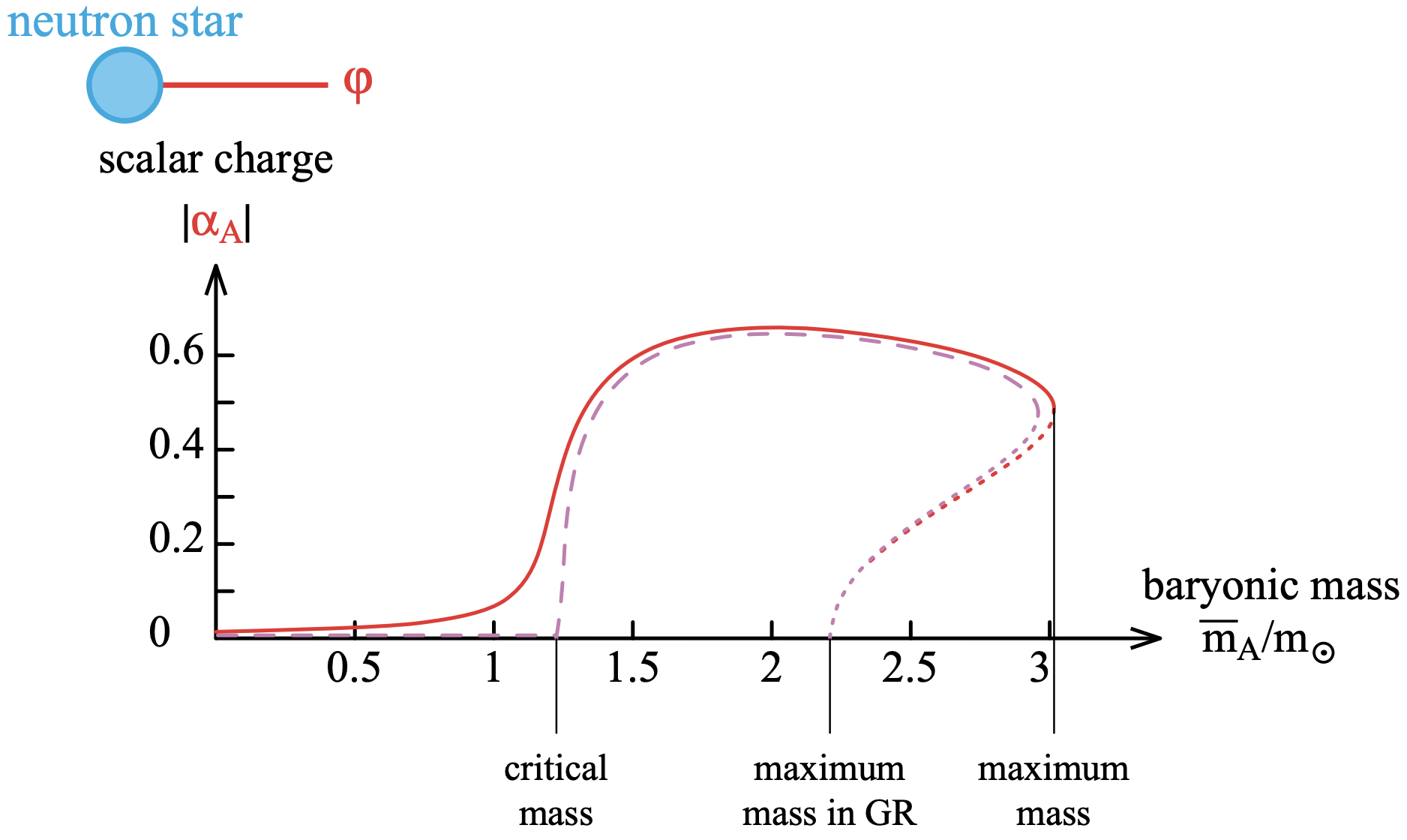}
\captionof{figure}{Dependence of the scalar charge on the baryonic mass to solar mass ratio for a compact object. The solid line represents the theory $T(-0.014,-6)$, the dashed line $T(0,-6)$ and the dotted line is the unstable configuration of the star. Figure taken from Damour 2009 \cite{damour2009binary}. \label{fig_alpha_mass}}
\end{minipage}
\end{figure}

We can now describe the concept of spontaneous scalarization. As explained by Damour and Esposito-Farèse 1996 \cite{damour1996tensor}, for small enough $\beta_0$, a compact object, such as neutron star, with a mass $m_A$ higher than a critical mass $m_c(\beta_0)$ depending on $\beta_0$, the coupling constant $\beta_A(\varphi_0)$ becomes of order of unity. On figure \ref{fig_beta_mass} we draw the dependence of $m_c$ on $\beta_0$ for the model $A(\varphi)=\exp(\half \beta_0\varphi^2)$. We can see that for small $\beta_0$ the critical mass becomes of the order of the solar mass $m_{\odot}$. Furthermore, as shown in figure \ref{fig_alpha_mass} for a specific scalar-tensor theory $T(-0.014,-6)$, the coupling strength or scalar charge $\alpha_A$ becomes of the order of unity even for small $\alpha_0$ and one would observe such effect on the motion of pulsars. It is commonly accepted that the minimum value for $\beta_0$ that fits the current observations is $\beta_0\geq -5$ \cite{damour1996tensor}, which sets yet an other constraint on the possible scalar-tensor theories. 

\begin{figure}
    \centering
    \includegraphics[width=0.9\textwidth]{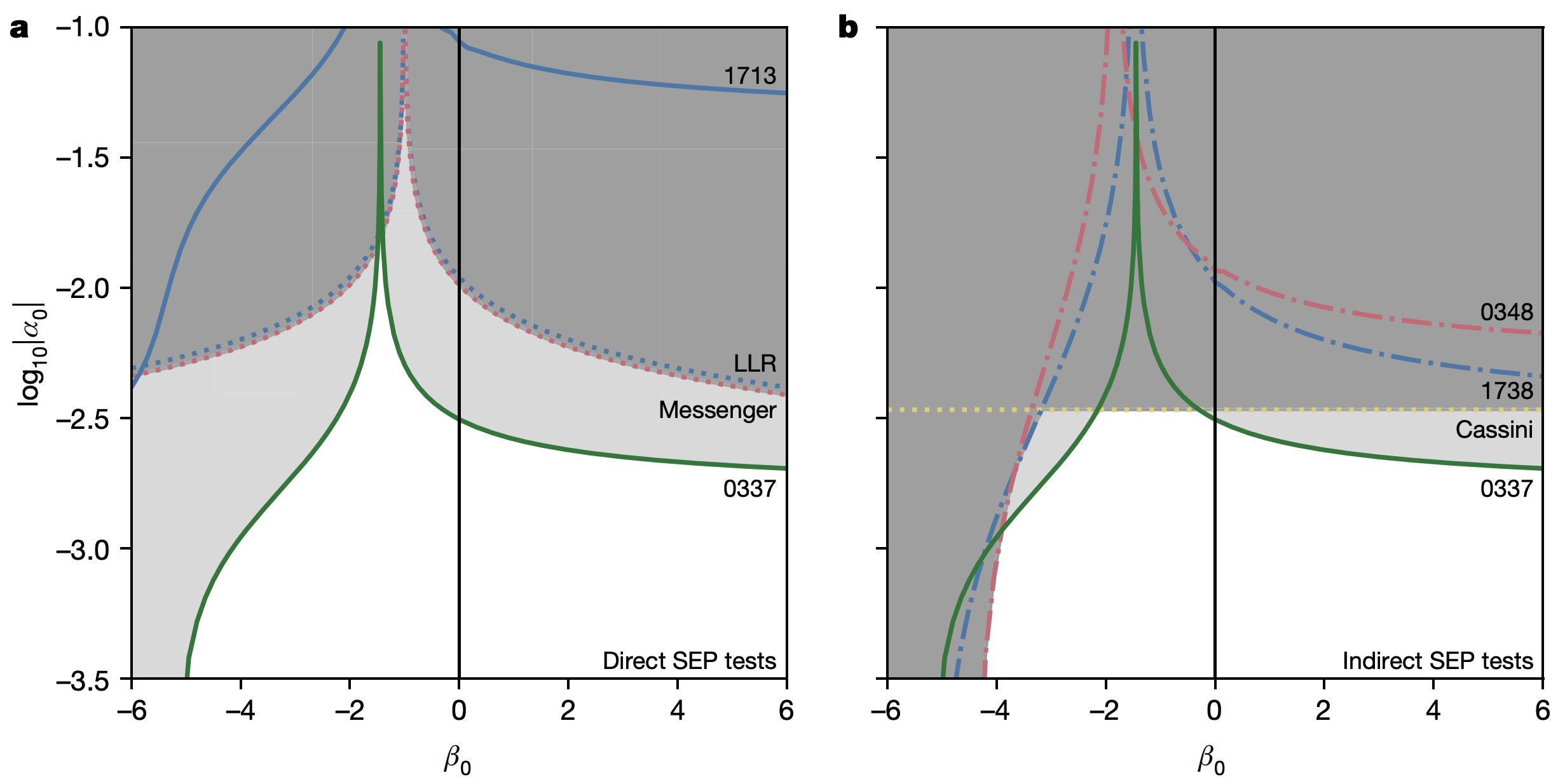}
    \captionof{figure}{Constraint on the asymptotic coupling constants $\alpha_0$ and $\beta_0$ of scalar-tensor theories for different tests using pulsars and solar system. The line labeled 0337 is the constraint using the triple system PSR J0337+1715. Figure taken from Archibald et al 2018 \cite{archibald2018universality}. \label{fig_archBound}}
\end{figure}

Using the constraint due to the limit on the Nordtvedt parameter represented in figure \ref{fig_a0b0constraint} and this new limit on $\beta_0$, we can refine the upper bound for the function $\alpha_0(\beta_0)$. This investigation was conducted by Archibald et al 2018 \cite{archibald2018universality} and their result is exposed in figure \ref{fig_archBound}. In the figure, the authors also added the previous most constraining tests and we can see that for a large range of values for $\beta_0$, the triple system test sets the currently most constraining limit. Furthermore contrary to figure \ref{fig_a0b0constraint} we can see the maximum limit of $\alpha_0$. 

\subsection{PPK Parameters for Scalar-Tensor Theories}

In this section, our goal is to write the PPK parameters for a binary system in a general scalar-tensor theory. We are now considering strong field effect, thus the asymptotic values $\alpha_0$ and $\beta_0$ are not sufficient anymore, we need to consider the effective coupling strength and its derivative. Both objects have effective coupling strength $\alpha_A$ and $\alpha_B$ together with their derivative $\beta_A$ and $\beta_B$, which represent the interaction of the different objects with the scalar field. To get to the expression for the PPK parameters, one can start with the action \eqref{action3} for the point mass objects approximation. Such a derivation was conducted by Horbatsch and Burgess 2012 \cite{horbatsch2012model} and we now give their results
\begin{subequations}
\label{eq_PPKST}
\begin{align}
    G_{AB}&=\tilde{G}(1+\alpha_A\alpha_B)\\
    \dot{\omega}&=\frac{n}{1-e^2}\left(\frac{G_{AB}Mn}{c^3}\right)^{2/3}\left(\frac{3-\alpha_A\alpha_B}{1+\alpha_A\alpha_B}-\frac{X_A\beta_B\alpha_A^2+X_B\beta_A\alpha_B^2}{2(1+\alpha_A\alpha_B)^2}\right)\\
    \gamma'&=\frac{eX_B}{n(1+\alpha_A\alpha_B)}\left(\frac{G_{AB}Mn}{c^3}\right)^{2/3}(X_B(1+\alpha_A\alpha_B)+1+k_A\alpha_B)\\
    r&=\tilde{G}(1+\alpha_0\alpha_B)\frac{m_B}{c^3}\\
    s&=\frac{nx_A}{X_B}\left(\frac{G_{AB}Mn}{c^3}\right)^{-1/3}\\
    \dot{P}_b&=\dot{P}_b^{mon}+\dot{P}_b^{dip}+\dot{P}_b^{qua}+\dot{P}_b^{kin}+\dot{P}_b^{gal},\label{eq_pdotST}
\end{align}
\end{subequations}
where $M=m_A+m_B$, $n=2\pi/P_b$, $X_i=m_i/M$ and $x_A$ is the projected semi-major axis of the object $A$. The exponents in equation \eqref{eq_pdotST} account for monopole, dipole, quadripole, kinetic and galaxy respectively. The expressions for the different contributions to $\dot{P}_b$ are long and exposed in Horbatsch and Burgess 2012 \cite{horbatsch2012model} if needed, we will only comment on their interpretation here. $\dot{P}_b^{mon}$, $\dot{P}_b^{dip}$ and $\dot{P}_b^{qua}$ are the contributions due to monopole, dipole and quadripole radiations. Then $\dot{P}_b^{kin}$ accounts for the contribution due to the relative speed between the system and the solar system. Finally $\dot{P}_b^{gal}$ describes the galactic contribution due to the relative acceleration between the system and the solar system. We can also relate the PPK parameters of scalar-tensor theory with the general PPK parameters \eqref{EqPPK}, that we derived earlier. 

Looking at the equations \eqref{eq_PPKST}, we can first observe the direct effect of spontaneous scalarization on the motion of binary systems, through the effective coupling strength $\alpha_A$. Contrary to the GR case, we have 7 unknown parameters
$\{m_A,m_B,\alpha_A,\alpha_B,\beta_A,\beta_B,k_A\}$ instead of 2, thus an analysis such as the one we represented in figure \ref{MMdiagram}, is not straight forward anymore. Note that in order to recover GR far from the binary system, we set $\alpha_0=0$, which leads to $\tilde{G}=G$ and $r=m_BG/c^3$. However there is a way that we could completely constrain scalar-tensor theories. In the case of a double pulsar, one could observe the PPK parameters for both objects and thus have 2 sets of equations \eqref{eq_PPKST}, where we exchange $A\leftrightarrow B$. Since the equations for $\pdot_b$ are $\gamma'$ are symmetric under $A\leftrightarrow B$, we have 8 independent equations. Furthermore we can also observe the semi-major axis ratio between the two pulsars to raise the number of independent equations to 9. In order to describe a double pulsar system with scalar-tensor theories, one need 8 different parameters $\{m_A,m_B,\alpha_A,\alpha_B,\beta_A,\beta_B,k_A,k_B\}$, thus it is possible to completely constrain these theories using double pulsar timing. Unfortunately the only know double pulsar system PSR  J0737-3039A/B, is not suitable for such an observation. Indeed the parameters $r$ and $s$ were not observable for one of the pulsar, because of its orientation with Earth and furthermore it has gone silent since 2008 \cite{lyne2006review, burgay2012double}.

%---------------------------

\section{Massive Gravity} \label{section_MG}
The idea to add a mass to the field propagating gravity has been around for several decades and many such theories has emerged in the 20th century, however they all suffered from what is called Boulware-Deser ghosts \cite{de2011resummation}. In a relatively recent paper, de Rham et al 2011 \cite{de2011resummation} were able to construct a ghost free massive gravity theory and revived the interest in such theories. In this section, we want to construct the ghost-free theory of massive gravity and understand the implications of such a modification of general relativity on binary systems, through the Vainshtein mechanism.
\subsection{Ghost-Free Construction of Massive Gravity}
In order to better understand the concept of massive gravity, we shortly construct the ghost-free theory using the work of Heisenberg 2019 \cite{heisenberg2019systematic}. For a more detailed derivation of massive gravity we refer to de Rham 2014 \cite{de2014massive}. As a starting point, we consider the Lagrangian of the linear theory of Fierz-Pauli
\begin{align}
    \LL = -h^\ab \EE^{\munu}_\ab h_\munu,\label{eq_FPL}
\end{align}
where
\begin{align}
   \EE^{\munu}_\ab h_\munu=-\half(\Box h_\ab -2\partial_{\mu}\partial_{(\alpha}h_{\beta)}^{\mu}+\partial_\alpha \partial_{\beta}h-\eta_\ab(\Box h-\partial_\mu \partial_{\nu}h^\munu)).
\end{align}
Adding a mass term to the Lagrangian \eqref{eq_FPL} will break the gauge invariance and will give a priori 6 propagating degrees of freedom to the tensor field. However a massive spin-2 field should only propagate 5 degrees of freedom, two helicity-2, two helicity-1 and one helicity-0. The solution to this problem is to insert an appropriate mass term, which decouples the ghost degree of freedom by giving it an infinite mass. Adding such a mass term and a matter contribution gives
\begin{align}
    \LL=-h^\ab \EE^{\munu}_\ab h_\munu-\frac{m^2}{2}(h_\munu h^\munu-h^\mu_\mu h^\nu_\nu)+\frac{1}{M_P}h_\munu T^\munu,
\end{align}
where $m$ is the mass of the graviton and $M_P$ is the Planck mass. One can now try to understand the implications of such a modification of the linearized General Relativity. We can first represent the extra degrees of freedom using a vector field $A_\mu$ for helicity-1 and a scalar field $\pi$ for helicity-0, using the transformations $h_\munu\rightarrow h_\munu+\partial_\mu A_\nu+\partial_\nu A_\mu$ and $A_\mu\rightarrow A_\mu +\partial_\mu\pi$. Furthermore we can canonically normalize the fields $A_\mu\rightarrow \frac{1}{m}A_\mu$, $\pi\rightarrow\frac{1}{m^2}\pi$ and take the limit $m\rightarrow 0$ to get 
\begin{align}
    \LL=-h^\ab \EE^{\munu}_\ab h_\munu-\half F_\munu F^\munu +\frac{1}{M_P}h_\munu T^\munu -2(h_\munu \partial^\mu\partial^\nu\pi-h \Box \pi),
\end{align}
where $F_\munu=\partial_\mu A_\nu-\partial_\nu A_\mu$. We can see that the vector field decouples completely, whereas the scalar field interaction survives, thus even in the limit of a vanishing mass, the massive spin-2 field will propagate a helicity-0 degree of freedom. This phenomenon is at the origin of what is called vDVZ discontinuity. In short, contrary to the massive vector field case, in the limit of vanishing mass, the helicity-0 state does not decouple from the helicity-2 states. Such a consequence would have a noticeable effect on astrophysical observations and would invalidate the theory, however this phenomenon is only possible in the linearized theory. Indeed we will see later on that in the case of the non-linear theory, the Vainshtein mechanism solves this problem.

In order to construct a general theory of massive gravity, we need to include non-linear interactions and in particular the self-interactions of the tensor field. we first remark that up to now, we used the Minkowski metric to raise and lower indices, however one can extend our result to an arbitrary reference metric $\eta_\munu\rightarrow f_\munu$. Furthermore one could make this new reference metric dynamical, which is the realm of bigravity theories. In order to keep only five degrees of freedom in our theory we can construct a general Lagrangian using the elementary symmetric polynomials with the fundamental matrix $\KK^\mu_\nu=\delta_\nu^\mu-(\sqrt{g^{-1}f})^\mu_\nu$, where $g^{-1}f$ is the matrix multiplication between the two metric. The resulting action for a general ghost-free massive gravity is 
\begin{align}
    S=\int\dd^4x\left[\frac{M_P^2}{2}\sqrt{-g}\left(R[g]-\frac{m^2}{2}\sum\limits_{n=1}^4\alpha_nU_n[\KK]\right)+\LL_{matter}\right],
\end{align}
where $R[g]$ is the Ricci scalar for $g_\munu$, $\alpha_n$ are arbitrary parameters and $U_n[\KK]$ are the four first orders of the elementary sysmmetric polynomials, see Heisenberg 2019 \cite{heisenberg2019systematic}. As we have seen for the linearized case, for the limit $m\rightarrow 0$, the helicity-1 states decouple, thus we can only investigate the effects of the scalar field. In the case $h=A=0$, it is possible to write $\KK_\nu^\mu=\frac{1}{M_Pm^2}\Pi_\nu^\mu$, where $\Pi_\munu=\partial_\mu\partial_\nu \pi$ \cite{de2014massive}. As we know from observations, the Planck mass is very large and the mass of the graviton is predicted to be of the order $m\sim H_0\sim 10^{-33}$eV, where $H_0$ is the Hubble constant, thus it is natural to consider the limit for $m\rightarrow 0$, $M_P\rightarrow \infty$, such that $\Lambda=M_Pm^2$ is constant. In this limit, called the decoupling limit, the Einstein-Hilbert term reduces to its linearized version and the ghost-free massive gravity is given by
\begin{align}
    \LL=-\frac{1}{4}h^\munu\EE^\ab_\munu h_\ab+h^\munu\sum\limits_{n=1}^3\frac{a_n}{\Lambda^{3(n-1)}}X_\munu^{(n)},\label{eq_lagrangian-massiveG}
\end{align}
where 
\begin{subequations}
\begin{align}
    X^{(1)}_\munu&=\epsilon_{\mu}^{\ \alpha\rho\sigma}\epsilon_\nu^{\ \beta}{}_{\rho\sigma}\Pi_\ab,\\
    X^{(2)}_\munu&=\epsilon_{\mu}^{\ \alpha\rho\gamma}\epsilon_\nu^{\ \beta\sigma}{}_\gamma\Pi_\ab\Pi_{\rho\sigma},\\
    X^{(3)}_\munu&=\epsilon_{\mu}^{\ \alpha\rho\gamma}\epsilon_\nu^{\ \beta\sigma\delta}\Pi_\ab\Pi_{\rho\sigma}\Pi_{\gamma\delta},
\end{align}
\end{subequations}
and $a_n$ are new constants related to $\alpha_n$. 

As mentioned earlier, the vDVZ patalogy is cured in the non-linear case by the Vainshtein mechanism. First introduced by Arkady Vainshtein, this mechanism says that for distances to the source $r$ smaller than a critical distance $r_V$, called Vainshtein radius, the linear model breaks down. Because of this, the effect of the helicity-0 state is screened for small distances, but still have cosmological effects on large scales. As an example we take the case of the solar system, the Vainshtein radius of the Sun is $r_V\sim 10^4$ pc \cite{de2013vainshtein}, thus the scalar degree of freedom has no effect on the solar system behaviour, as required. In the next part we will describe the Vainshtein mechanism more in detail in the case of a binary system and we will infer possible tests one can do for massive gravity.

\subsection{Vainshtein Mechanism for Binary Systems using Cubic Galileon}
In order to correctly describe the Vainshtein mechanism in the case of binary pulsars, we strongly rely on the work of de Rham et al 2013 \cite{de2013vainshtein}. Furthermore, for simplicity, we consider the simplest case where the Vainshtein mechanism plays a role, i.e. the cubic Galileon model. It is also enough in our case to work in the decoupling limit, for which the action of the model reads
\begin{align}
    S=\int\dd^4x \left[-\frac{1}{4}h^\munu\EE^\ab_\munu h_\ab-\frac{3}{4}(\partial\pi)^2\left(1+\frac{1}{3\Lambda^3}\Box\pi\right)+\frac{1}{2M_P}h^\munu T_\munu+\frac{1}{2M_P}\pi T\right].\label{eq_action-cubicGal}
\end{align}
One can see that the non-matter part of the action \eqref{eq_action-cubicGal} is very similar to the Lagrangian \eqref{eq_lagrangian-massiveG} of the decoupling limit of massive gravity up the order $\frac{1}{\Lambda^3}$, thus the cubic Galileon model gives us a good interpretation of the behaviour of massive gravity. Note that $\Lambda\sim 10^{-13}$eV and higher order terms in \eqref{eq_lagrangian-massiveG} have a stronger factor, which means that at low enough energies $E\lesssim\Lambda^3$, higher order contributions do not play a role. 

Our goal is to find the power $P$ of the gravitational radiation in the cubic Galileon model. We first assume a point mass object $T_0^\mu{}_\nu=-m\delta(\vb{x})\delta^\mu_0\delta^0_\nu$ and solving the equations of motion for $E(r)=\abs*{\grad \pi}$, we find
\begin{align}
    E_\pm(r)=\frac{\Lambda^3}{4r}\left(\pm\sqrt{9r^4+\frac{32r^3r_V}{\pi}}-3r^2\right),\quad\textrm{where}\quad r_V=\frac{1}{\Lambda}\left(\frac{m}{16M_P} \right)^{1/3}.
\end{align}
Since $E_+(r)\rightarrow 0$ when $r\rightarrow \infty$ and carries no ghost-like infinities, we choose this solution. From now on, we omit the subscript of $E_+$, i.e. we write $E$ instead of $E_+$. In order to compute the solution for a binary system, we use perturbation around the point mass source $T^\munu=T_0^\munu+\delta T^\munu$, where
\begin{align}
    \delta T^\mu_\nu=-\left(\sum\limits_{i=1,2}m_i\delta(\vb{x}-\vb{x}_i)-m\delta(\vb{x})\right)\delta^\mu_0\delta_\nu^0,
\end{align}
and $m_i$ are the objects mass and $m=m_1+m_2$. Since we split the source into a background part and a perturbative part, we can do the same for the scalar field, namely $\pi:=\pi_0+\sqrt{2/3}\phi$. The perturbation $\phi$ can be described with the Lagrangian 
\begin{align}
    \LL_\phi=&\half\lp 1+\frac{1}{2\Lambda^3}\lp E'+\frac{2E}{r}\rp\rp\dot{\phi}^2-\half\lp 1+\frac{4}{3\Lambda^3}\frac{E}{r}\rp(\partial_r\phi)^2-\label{eq_lagrangianphi}\\
    &-\half\lp 1+\frac{1}{2\Lambda^3}\lp E'+\frac{2E}{r}\rp\rp (\grad_\Omega\phi)^2+\frac{\phi}{\sqrt{6}M_P}\delta T, \nonumber
\end{align}
using spherical coordinates, where $\delta T=\delta T^\mu_\mu$. In order to find the power $P$ we use the effective action approach. One can find the radiation power with the effective action $S_{eff}$ using 
\begin{align}
    P=\int_0^\infty\dd\omega \omega f(\omega),\quad \textrm{where}\quad \int_0^\infty\dd \omega f(\omega)=\frac{2\textrm{Im}(S_{eff})}{P_b},
\end{align}
with the effective action calculated over one period. In our case, using the Lagrangian \eqref{eq_lagrangianphi}, we can write the effective action as
\begin{align}
    S_{eff}=&\int\dd^4x\LL_m+\frac{i}{12M_P^2}\int\dd^4x\dd^4x'\delta T(x)G_F(x,x')\delta T(x')+\\&+\textrm{helicity-2 contribution of GR},\nonumber
\end{align}
where $G_F(x,x')$ is the Green function associated to the d'Alembertian operator coming from \eqref{eq_lagrangianphi} (see de Rham et al 2013 \cite{de2013vainshtein} for a detailed expression). Since we are only interested in the imaginary part of $S_{eff}$, we only need to find the Green function. It is possible to describe $G_F$ using the Wightman functions
\begin{align}
    G_F(x,x')=\theta(t-t')W^+(x,x')+\theta(t'-t)W^-(x,x'),
\end{align}
where
\begin{align}
    W^+(x,x')=\sum\limits_{l,m}\int^\infty_0\dd\omega u^{\ }_{lm}(r,\Omega)u^*_{lm}(r',\Omega')e^{-i\omega (t-t')},
\end{align}
with $u^{\ }_{lm}(r,\Omega)e^{-i\omega t}$ are a complete set of mode functions solving the homogeneous equations of motion from \eqref{eq_lagrangianphi}. Using the prior results, it is possible to solve for $f(\omega)$ and finally one can obtain the radiated power 
\begin{align}
    P=\frac{\pi}{3M_P^2}\sum\limits_{n=0}^\infty\sum\limits_{l,m}n\Omega_P\abs{M_{lmn}}^2,\label{eq_power}
\end{align}
where $\Omega_P=2\pi/P_b$ and 
\begin{align}
    M_{lmn}=\frac{1}{P_b}\int_0^{P_b}\dd t\dd^3x\;u^{\ }_{lm}(r,\Omega)e^{-in\Omega_Pt}\delta T.
\end{align}
In order to completely describe the emitted power, we need to specify the mode functions $u_{lm}$. First we can decompose them as $u^{\ }_{lm}(r,\Omega)=u_l(r)Y_{lm}(\Omega)$, where $Y_{lm}$ are the spherical harmonics. Then one can write $u_l$ in terms of $r$ and the emitted angular frequency $\omega\sim\Omega_P$
\begin{align}
    u_l=
    \begin{cases}
    \lp\frac{3}{8\pi r_V^3r\omega^2}\rp^{1/4}\cos\lp\frac{\sqrt{3}}{2}\omega r\rp,\quad \textrm{for}\quad \omega^{-1}\ll r\ll r_V\\
    \frac{1}{r\sqrt{\pi\omega}}\cos(\omega r),\quad\textrm{for}\quad r\gg r_V.
    \end{cases}
\end{align}
We now write the power in terms of the multipole expansion, i.e. $P\cong P_{M}+P_D+P_Q$. Using the equation for the emitted power \eqref{eq_power} and the Keplerian orbit approximation
\begin{align}
    r_{1,2}(t)=\frac{a_1(1-e^2)}{1+e\cos(\Omega_P t)}\frac{m_{2,1}}{m},
\end{align}
with $e$ the eccentricity, it is possible to compute the first orders of the multipole expansion. We now write the result for the monopole, dipole and quadripole radiation, for a detailed derivation of the result see de Rham et al 2013 \cite{de2013vainshtein}
\begin{subequations}
\label{eq_multipole_power}
\begin{align}
    P_M&=\frac{25\pi}{3}\frac{b^2}{16}\frac{(\Omega_Pa_1)^4}{(\Omega_Pr_V)^{3/2}}\frac{M^2_M}{M^2_P}\Omega_P^2\sum\limits_{n>0}\abs{I_n^M(e)}^2,\label{eq_power_monopole}\\
    P_D&=\frac{c_1^2}{8}\frac{(\Omega_Pa_1)^6}{(\Omega_Pr_V)^{3/2}}\frac{M^2_D}{M^2_P}\Omega_P^2\sum\limits_{n\geq 0}\abs{I_n^D(e)}^2,\\
    P_Q&=\frac{5\lambda^2}{32}\frac{(\Omega_Pa_1)^3}{(\Omega_Pr_V)^{3/2}}\frac{M^2_Q}{M^2_P}\Omega_P^2\sum\limits_{n\geq 0}\abs{I_n^Q(e)}^2,\label{eq_power_quadripole}
\end{align}
where $b\cong 0.69$, $c_1\cong 0.4$, $\lambda\cong0.21$ (exact expressions in \cite{de2013vainshtein}), $M_M=(m_1m_2^2+m_2m_1^2)^2/m^2$, $M_D=m_1m_2(m_1^2-m_2^2)/m^3$, $M_Q=m_1m_2(m_1^{1/2}+m_2^{1/2})/m^{3/2}$ and 
\begin{align*}
    I_n^M(e)&=n^{9/4}\sqrt{1-e^2}e^{-n}(\sqrt{1-e^2}-1)^n(1+n\sqrt{1-e^2}),\\
    I_n^D(e)&=(1-e^2)^3\frac{n^{13/4}}{2\pi}\int_0^{2\pi}\frac{\exp(-i(n-1)x)}{1+e\cos x}\dd x,\\
    I_n^Q(e)&=(1-e^2)^{3/2}\frac{n^{7/4}}{2\pi}\int_0^{2\pi}\frac{\exp(-i(n-2)x)}{(1+e\cos x)^{3/2}}\dd x.
\end{align*}
\end{subequations}
After all these derivations, one might wonder where the Vainshtein mechanism enters into play. Indeed we have yet to discuss the role of the mechanism in the gravitational radiation. As a first hint, we can observe that the Vainshtein radius appears in the multipole expansion \eqref{eq_multipole_power}. The term $(\Omega_Pr_V)^{-3/2}$ acts as a suppression factor in the monopole formula \ref{eq_power_monopole}, which reduces its order of magnitude. In the case of the quadripole formula \ref{eq_power_quadripole}, we can directly compare with the quadripole radiation of General Relativity and we obtain
\begin{align}
    \frac{\left.P_Q\right|_\pi}{\left.P_Q\right|_{GR}}=q(\Omega_Pr_V)^{-3/2}(\Omega_Pa_1)^{-1},
\end{align}
where $q$ depends on the eccentricity and the mass difference of the binary system. In the case of the Pulse-Taylor pulsar, $q\cong 0.08$ \cite{de2013vainshtein}. Here again one can see that the Vainshtein mechanism acts as a suppression factor and reduces the magnitude of the scalar field radiation. Concerning the dipole contribution, it is also suppressed with a similar factor and it is usually several order of magnitude lower than the monopole radiation. If one would do the same derivation in the case of the quartic Galileon model, closer to the complete theory of massive gravity in the decoupling limit \eqref{eq_lagrangian-massiveG}, we would obtain a suppression factor $(\Omega_Pr_V)^{-2}$ \cite{de2013galileon}.

Our goal is now to find a way to test the cubic Galileon model or by similarity to test massive gravity through the Vainshtein mechanism and the radiation power using pulsar timing. It is fairly straight forward to connect the loss of energy due to gravitational emission with the decrease in orbital period. In the non-relativistic case we have the relation
\begin{align}
    \pdot_b=-\frac{3}{2}\frac{P_b}{E_b}P,
\end{align}
where $E_b$ is the energy of the binary system. To better understand the amplitude of the effect of the scalar field on the gravitational radiation of a binary pulsar, we take the Hulse-Taylor system as an example. The comparison between the contribution of the scalar field and the contribution of General relativity is exposed in table \ref{table_Pb_massive_gravity}. One can see that the effect of the scalar field is several orders of magnitude lower than the effect of General Relativity and the measurements are not sensitive enough to validate or invalidate the theory, in the case of a graviton mass $m=1.54\cdot 10^{-33}$ eV. If one now work in reverse, i.e. use observations in order to find a limit for the mass of the graviton, we can set an upper bound for the mass of the graviton using pulsar timing. Current measurements of pulsar binary systems set the lowest upper bound for the mass of the graviton at \cite{de2013vainshtein}
\begin{align}
    m<10^{-27},
\end{align}
which is 6 orders of magnitude higher than the required precision to properly test massive gravity. However pulsar timing measurements do not set the most constraining limit on the graviton mass. Indeed solar system measurements set the lowest upper bound to $m<10^{-32}$ \cite{seymour2018testing}, which gives high hope that we will soon be able to properly test massive gravity. The recent observation of the neutron star merger in 2017 was also able to set an upper bound on the mass of the graviton, however the limit is 10 orders of magnitude lower than the solar system experiments, i.e. $m<10^{-22}$ \cite{baker2017strong}.

\begin{table}[]
\centering
\begin{tabular}{|c|c|c|c|c|}
\hline
$\pdot_b$ monopole  & $\pdot_b$ dipole & $\pdot_b$ quadripole & $\pdot_b$ GR        & $\sigma$            \\ \hline
$4.5\cdot 10^{-22}$ & $\sim 10^{-30}$  & $2.0\cdot 10^{-22}$  & $2.4\cdot 10^{-12}$ & $5.1\cdot 10^{-15}$ \\ \hline
\end{tabular}
\captionof{table}{Comparison between the scalar field contributions to the decrease in orbital period in the cubic Galileon framework with the General Relativity case. We took $m=1.54\cdot 10^{-33}$ eV for the mass of the graviton. The experimental uncertainty $\sigma$ is around the value for GR. (from de Rham et al 2013 \cite{de2013vainshtein}). \label{table_Pb_massive_gravity}}
\end{table}

%--------------------------- 

\section{Conclusion}
After a brief introduction on pulsar's properties and pulsar timing techniques, we described the most general weak field gravity theory using the post-Newtonian formalism together with the PPN parameters. In particular we mentioned all the parameters with their interpretation and predicted value for General Relativity and scalar-tensor theories. Later on we constructed a general timing formula using the post-Keplerian formalism and the PPK parameters. We also mentioned each parameter and gave their prediction for General Relativity. Thanks to these PPK parameters we were able to construct a powerful test of General Relativity, namely the mass-mass diagram, and using the observations of the double pulsar PSR J0737-3039A/B, we could measure the mass of the two pulsars according to GR and show that the theory is consistent in this framework. We only showed the most constraining diagram in this review, one could do this test for any binary systems, however the fact that J0737-3039A/B is a double pulsar system gave us a better measurements. Since this system is the only one of its kind we have yet observed, it is hard to construct a more restrictive diagram than the one presented in figure \ref{MMdiagram}. The only hope is to find an other double pulsar system, which could happen with the use of the new radio telescope FAST. 

After introducing the relevant theoretical basis needed for the report, we could deal with the tests of the strong equivalence principle. We first associated the parameters $\alphah_1$ and $\alphah_2$ to a violation of the Lorentz invariance. We then constrained the first parameter up to the order $\alphah_1\lesssim 10^{-5}$ by measuring the eccentricity vector of the binary system PSR J1738+0333. Because a non-zero value for $\alphah_2$ would induce a spin precession of a solitary pulsar, we could set a bound $\alphah_2\lesssim 10^{-9}$ with observations of the pulsars PSR B1937+21 and PSR J1744-1134, which have a very short spin period. Using the method of the eccentricity vector, we were also able to set a bound on the parameter representing a violation of Lorentz invariance and energy conservation, i.e. $\alphah_3\lesssim 10^{-20}$. In order to test for position invariance, we used the parameter $\xih$, which we could constrain up to the order $\xih\lesssim 10^{-9}$, with the spin precession method. An other test we did for the position invariance is the variation of Newton's constant, we were able to set a bound by observations of the variation of orbital period for the binary system PSRJ1012+5307 and we set the upper limit to $\gdot/G\lesssim 10^{-12}$. This limit is not the most constraining bound for the variation of Newton's constant, indeed solar system tests reaches one order of magnitude lower. Finally we tested the gravitational weak equivalence principle with the Nordtvedt parameter using the surprising triple system PSR J0337+1715. By measuring the acceleration of each object, one can set a constrain up to order $\etah\lesssim 10^{-5}$. The limits on the PPN parameters set by pulsar timing techniques are stronger than the ones set by solar system experiments. However since pulsars are strong self-graviting bodies and the post-Newtonian formalism is a weak field approximation, the PPN parameters are not completely satisfactory to describe the motion of such objects. One then need to be careful when comparing solar system experiments and pulsar observations. We can however reasonably argue that strong field corrections to the PPN parameters in any theories, would increase their deviation from zero, thus a limit on a strong field PPN parameter can induce a similar limit on its weak field counter part. The beginning of the survey by the new radio telescope FAST will boost the discoveries of new pulsar systems and one can hope that more appropriate systems to test PPN parameters will be discovered. In particular, the discovery of a new double pulsar system would allow for better mass-mass diagram and one could potentially test for all the parameters of scalar-tensor theories. Indeed, one need 8 different parameters to fully describe the behaviour of a double pulsar system and one could observe 9 independent post-Keplerian parameters, thus it is possible to fully constrain scalar-tensor theories with pulsar timing techniques. Furthermore, if one finds a black hole-pulsar system, it would be possible to constrain the variation of orbital period by a few orders of magnitude \cite{seymour2018testing}.

In order to continue the analysis of the possible constraints one can set using pulsars, we constructed a general scalar-tensor theory and derived the PPN and PPK parameters for such a general theory. In particular we showed that, in the weak field limit, we can describe all the scalar-tensor theories with 2 parameters $\alpha_0$ and $\beta_0$. We could directly deduce a bound on the possible scalar-tensor theories using the limit on the Nordtvedt parameter $\etah$. The result of this study is exposed in figure \ref{fig_a0b0constraint}. Furthermore using the effect of spontaneous scalarization of pulsars, we could set a lower limit on $\beta_0$ and combining both bounds we could understand the figure \ref{fig_archBound}, which represents the set of surviving scalar-tensor theories. The derivation of the PPK parameters showed us that we need 7 parameters to describe the 5 post-Keplerian parameters and thus it is complicated to find a bound on possible scalar-tensor theories. However we also mentioned that in the case of a double pulsar system, we only need 8 parameters of the theory to describe 9 independent PPK parameters, thus it would be possible to constrain the set of possible scalar-tensor theories. Unfortunately one of the pulsars of the only know double pulsar system is not detectable anymore and some of the required PPK parameters for this silent pulsar cannot be observed. 

Finally we looked at the case of massive gravity. We first constructed the ghost-free theory and then we discussed the decoupling limit where the mass of the graviton is very small and the Planck mass is very large. In this limit, the theory is very similar to the simpler theory of the quartic Galileon, for which one can predict new behaviours of binary system. Analyzing the quartic Galileon theory is outside the scope of this review, so in order to understand the implications of the theory we discussed the simpler cubic Galileon theory. In this framework we described the Vainshtein mechanism, which acts as a screening mechanism of the scalar mode at short distances and we could set an upper bound on the mass of the graviton to $m\lesssim 10^{-27}$. The predicted mass of the graviton to account for cosmological observations is of the order $m\sim 10^{-33}$, thus pulsar timing is not sufficient to probe at the required order of magnitude. The best hope is coming from solar system experiments, indeed these latter could already set an upper limit down to $m\lesssim 10^{-32}$.

 %---------------------------

\newpage
\bibliographystyle{plain}
\bibliography{ref}

\end{document}